%% file: SequentialCS.tex
\documentclass[10pt,onecolumn]{IEEEtran}
\usepackage[LGR,T1]{fontenc}
\usepackage[latin9]{inputenc}
\usepackage{xcolor}
\usepackage{amsmath}
\usepackage{amsthm}
\usepackage{amssymb}
\usepackage[unicode=true,
 bookmarks=true,bookmarksnumbered=true,bookmarksopen=true,bookmarksopenlevel=3,
 breaklinks=false,pdfborder={0 0 1},backref=false,colorlinks=true]
 {hyperref}
\hypersetup{
 pdfborderstyle=,pdfborderstyle={},pdfborderstyle={},pdfborderstyle={},pdfborderstyle={},pdfborderstyle={},pdfborderstyle={},pdfborderstyle={},pdfborderstyle={},pdfborderstyle={},pdfborderstyle={},pdfborderstyle={},pdfborderstyle={},pdfborderstyle={},pdfborderstyle={},pdfborderstyle={},pdfborderstyle={},pdfborderstyle={},pdfborderstyle={},pdfborderstyle={},pdfpagelayout=OneColumn,pdfnewwindow=true,pdfstartview=XYZ,plainpages=false,linkcolor=blue,urlcolor=blue,citecolor=red,anchorcolor=blue,linkcolor=blue,urlcolor=blue,citecolor=red,anchorcolor=blue}

\makeatletter
\theoremstyle{plain}
\newtheorem{thm}{\protect\theoremname}
\theoremstyle{remark}
\newtheorem{rem}{\protect\remarkname}
\theoremstyle{plain}
\newtheorem{lem}{\protect\lemmaname}

\usepackage{color}
\usepackage{cite}
\input{preamble}

\newcommand{\blue}{\color{blue}}

\newcommand{\black}{\color{black}}

  \providecommand{\lemmaname}{Lemma}
  \providecommand{\remarkname}{Remark}
\providecommand{\theoremname}{Theorem}

\allowdisplaybreaks[1]
\makeatother

\providecommand{\lemmaname}{Lemma}
\providecommand{\remarkname}{Remark}
\providecommand{\theoremname}{Theorem}

\begin{document}
\title{Sequential Channel Synthesis}
\author{Lei Yu and Venkat Anantharam\thanks{L. Yu is with the School of Statistics and Data Science, LPMC \& KLMDASR,
Nankai University, Tianjin 300071, China (e-mail: leiyu@nankai.edu.cn).
V. Anantharam is with the Department of Electrical Engineering and
Computer Sciences, University of California, Berkeley, CA 94720, USA
(e-mail: ananth@berkeley.edu). The authors were supported by the NSF
grants CNS--1527846, CCF--1618145, CCF-1901004, CIF-2007965, the
NSF Science \& Technology Center grant CCF--0939370 (Science of Information),
and the William and Flora Hewlett Foundation supported Center for
Long Term Cybersecurity at Berkeley. The first author was also supported
in part by the NSFC grant 62101286 and the Fundamental Research Funds
for the Central Universities of China (Nankai University).}}
\maketitle
\begin{abstract}
The channel synthesis problem has been widely investigated over the last
decade. In this paper, we consider the sequential version in which
the encoder and the decoder work in a sequential way. Under a mild
assumption on the target joint distribution we provide
a complete (single-letter) characterization of the solution for the point-to-point
case, which shows that the canonical symbol-by-symbol mapping is not
optimal in general, but is indeed optimal if we make some additional assumptions
on the encoder and decoder. 
%We also extend this result to the broadcast
%scenario, the interactive communication scenario, and the energy harvesting
%scenario. 
%\color{magenta}
We also extend this result to the broadcast
scenario and the interactive communication scenario.
%\color{black}
%We provide either a complete characterization of the solution or bounds for these
%settings. 
%\color{magenta}
We provide bounds in the broadcast setting and 
a complete characterization of the solution under a mild condition
on the target joint distribution in the interactive communication case.
%\color{black}
Our proofs 
%in this paper 
are based on a R\'enyi entropy
method. 
\end{abstract}

\section{\label{sec:Introduction}Introduction}

The study of the synthesis of distributions can be traced back to the seminal
work by Wyner \cite{Wyner}
%\color{magenta}
where the problem studied was to characterize the smallest rate, in bits per symbol,
at which common randomness needs to be provided to two agents, Alice and Bob, each having
an arbitrary amount of private randomness, such that each of them can separately generate
a sequence of random variables from respective finite sets, with the joint distribution being
close to that of an i.i.d. sequence with a desired joint distribution at each symbol time
(the notion of approximation in \cite{Wyner} is based on relative entropy). 
Wyner used this framework to
define a notion of the common information 
of two dependent sources (the ones being synthesized by Alice and Bob respectively), 
which is known nowadays as Wyner's common information.
\iffalse
in which Wyner used this framework to
define the common information (known as Wyner's common information
nowadays) of two correlated sources. 
\color{magenta}
blah \cite{Cuff10} blah \cite{Gohari} blah \cite{Borkar}
\color{black}
\fi
This formulation is of considerable interest for problem of distributed control and 
game theory with distributed agents \cite{Borkar} because of the need to 
randomize for strategic reasons. It was generalized to the context of
networks by Cuff et al. \cite{Cuff10} where, in particular, the formulation allows 
for communication between the agents attempting to create i.i.d. copies of a target 
joint distribution, with the communication occurring at the level of blocks
of symbols, see also \cite{Gohari}.
\iffalse
By choosing proper measures of
the approximation level,  Wyner's definition of common information
can be extended to a variant version in which we seek for the minimum
communication rate required for a pair of sender and receiver to synthesize
a channel in a distributed way. 
\fi
For instance, for two agents, one can seek to find the minimum
communication rate required for a pair of sender and receiver to synthesize
a channel with a given input distribution in a distributed way. 
%\color{black}
Specifically, the sender and receiver
share a sequence of common random variables $W^{n}$. After observing
a source $X^{n}\sim\pi_{X}^{n}$ and the common randomness $W^{n}$,
the sender generates bits and send them to the receiver, who
generates another source $Y^{n}$ according to the common randomness
$W^{n}$ and the bits that he/she receives. They cooperate in such
a way so that the channel induced by the code $P_{Y^{n}|X^{n}}$ is
close to a target channel $\pi_{Y|X}^{n}$. If the closeness here
is measured by the total variation (TV) distance between $\pi_{X}^{n}P_{Y^{n}|X^{n}}$
and the target joint distribution $\pi_{XY}^{n}$, this channel synthesis
problem was investigated in \cite{bennett2002entanglement,winter2002compression,Cuff,bennett2014quantum}
and the minimum communication rate was completely characterized by
Cuff \cite{Cuff}. 
The exact synthesis of 
%\color{magenta}
such
%\color{black}
a channel was considered
in \cite{harsha2010communication,Kumar,li2017distributed,yu2019exact}
%and the exact synthesis with unlimited shared randomness was also
%studied in \cite{bennett2002entanglement,bennett2014quantum}. The
%\color{magenta}
where
%\color{black}
exact synthesis here means that the synthesized channel $P_{Y^{n}|X^{n}}$
is exactly equal to the target channel $\pi_{Y|X}^{n}$. The characterization
of the minimum communication rate for exact synthesis (given the
shared randomness rate) is an interesting but hard problem. It is
still open until now except for some cases: the exact synthesis
for symmetric binary erasure source (completely characterized by Kumar,
Li, and El Gamal \cite{Kumar}) and the doubly symmetric binary source
(completely characterized by Yu and Tan \cite{yu2019exact}). 

In this paper, we consider an arguably more natural variant of the channel synthesis
problem, which we call the \emph{sequential channel synthesis problem}, in
which the encoder and the decoder work in a sequential way. 
%\color{magenta}
Under a mild assumption on the target joint distribution 
%\color{black}
we provide
a complete (single-letter) characterization for the point-to-point
case, which shows that the canonical symbol-by-symbol mapping is not
optimal in general (but we also show that it is indeed optimal if we make 
an additional assumption
on the encoder and decoder). We also extend this result to the broadcast
scenario and the interactive communication scenario, 
%and the energy harvesting scenario,
where we provide bounds in the 
former case and a complete solution
in the latter case 
under a mild assumption on the target joint distribution.
Our proofs in this paper are based on a R\'enyi entropy
method. 

\subsection{Problem Formulation }

Let $\mathcal{W}$, $\mathcal{X}$, $\mathcal{Y}$ and $\mathcal{B}$
be finite sets. %Consider the following sequential channel synthesis problem. 
Alice and Bob share a sequence of i.i.d. random variables $\left\{ W_{i}\right\} $
taking values in $\mathcal{W}$, with each $W_{i}\sim P_{W}$. Let
$\left\{ X_{i}\right\} $ be a sequence of i.i.d. random variables
taking values in $\mathcal{X}$, with each $X_{i}\sim\pi_{X}$. We
assume that $\left\{ X_{i}\right\} $ and $\left\{ W_{i}\right\} $
are independent. $\left\{ X_{i}\right\} $ is called the source sequence.

Consider the following sequential channel synthesis problem. At the
epoch $k$, upon observing the common random sequence\footnote{Throughout this paper, for any sequence %of i.i.d. random variables
%$\left\{ Z_{i}\right\} $, 
$(z_{k},k\ge1)$, we use the notation %$Z^{k}:=\left(Z_{1},Z_{2},...,Z_{k}\right)$
$z^{k}:=(z_{1},\ldots,z_{k})$ for $k\ge1$.} $W^{k}$, the source sequence $X^{k}$, and previous communication
random variables $B^{k-1}$, Alice generates $B_{k}\in\mathcal{B}$
by using a random mapping with conditional distribution $P_{B_{k}|W^{k}X^{k}B^{k-1}}$,
and then sends $B_{k}$ to Bob. At the epoch $k$, upon observing
$W^{k}$, $B^{k}$, and the previous outputs $Y^{k-1}$, Bob generates
$Y_{k}$ taking values in $\mathcal{Y}$, by using a random mapping
with conditional distribution $P_{Y_{k}|W^{k}B^{k}Y^{k-1}}$. Given
a target channel $\pi_{Y|X}$, the goal for Alice and Bob is to cooperate
in this sequential manner to minimize the Kullback-Leibler (KL) divergence
$D\left(P_{Y^{n}|X^{n}}\|\pi_{Y|X}^{n}|\pi_{X}^{n}\right)$ of the
synthesized joint distribution $\pi_{X}^{n}P_{Y^{n}|X^{n}}$ with
respect to the target joint distribution $\pi_{X}^{n}\pi_{Y|X}^{n}$,
where $\pi_{X}^{n}(x^{n}):=\prod_{i=1}^{n}\pi_{X}(x_{i})$ and $\pi_{Y|X}^{n}(y^{n}|x^{n}):=\prod_{i=1}^{n}\pi_{Y|X}(y_{i}|x_{i})$.
%(Since $\pi_{X}^{n},\pi_{Y|X}^{n}$ are product, sometimes, we also denote them respectively  as $\pi_{X}^{n},\pi_{Y|X}^{n}$.)
Here the conditional KL divergence for two conditional distributions
$P_{U|V}$ and $\pi_{U|V}$ conditioned on the marginal distribution
$\pi_{V}$ is defined as 
\[
D\left(P_{U|V}\|\pi_{U|V}|\pi_{V}\right):=D\left(P_{U|V}\pi_{V}\|\pi_{U|V}\pi_{V}\right).
\]
The channel synthesized by Alice and Bob can be expressed as 
\[
P_{Y^{n}|X^{n}}\left(y^{n}|x^{n}\right):=\sum_{b^{n}}\sum_{w^{n}}P_{W}^{n}\left(w^{n}\right)\prod_{k=1}^{n}P_{B_{k}|W^{k}X^{k}B^{k-1}}\left(b_{k}|w^{k},x^{k},b^{k-1}\right)\prod_{k=1}^{n}P_{Y_{k}|W^{k}B^{k}Y^{k-1}}\left(y_{k}|w^{k},b^{k},y^{k-1}\right),
\]
where $P_{W}^{n}\left(w^{n}\right):=\prod_{i=1}^{n}P_{W}(w_{i})$.
We are interested in characterizing 
\begin{equation}
\Gamma\left(\pi_{XY},P_{W}\right):=\lim_{n\to\infty}\frac{1}{n}\Gamma^{(n)}\left(\pi_{XY},P_{W}\right),\label{eq:-3}
\end{equation}
where 
\begin{equation}
\Gamma^{(n)}\left(\pi_{XY},P_{W}\right):=\inf_{\left\{ \left(P_{B_{k}|W^{k}X^{k}B^{k-1}},P_{Y_{k}|W^{k}B^{k}Y^{k-1}}\right)\right\} _{k=1}^{n}}D\left(P_{Y^{n}|X^{n}}\|\pi_{Y|X}^{n}|\pi_{X}^{n}\right),\label{eq:-31}
\end{equation}
and $\pi_{XY}(x,y):=\pi_{X}(x)\pi_{Y|X}(y|x)$. The limit in \eqref{eq:-3}
exists since $\Gamma^{(n)}\left(\pi_{XY},P_{W}\right)$ is subadditive
in $n$.

When $P_{W}$ is degenerate, i.e., $W_{i}$ is constant for all $i$,
then this corresponds to the case in which there is no common randomness. The optimal asymptotic KL divergence for this case is    denoted 
by $\Gamma_{0}\left(\pi_{XY}\right)$.

%\color{magenta}
We assume throughout that $|\mathcal{B}| \ge 2$, where $|\mathcal{B}|$ denotes the cardinality of $\mathcal{B}$,
since otherwise the problem is of no interest. 
%\color{black}
% The optimal asymptotic KL divergence for this case, which we denote
% by $\Gamma_{0}\left(\pi_{XY}\right)$, turns out to be equal to $\inf_{P_{W}}\Gamma\left(\pi_{XY},P_{W}\right)$.
% On the other hand, as the other extremal case, when Alice and Bob
% share an infinite amount of common randomness, the optimal asymptotic
% KL divergence is denoted  by $\Gamma_{\infty}\left(\pi_{XY}\right)$,
% which turns out to be equal to $\sup_{P_{W}}\Gamma\left(\pi_{XY},P_{W}\right)$.
\iffalse
\color{red}
VA question: Each of the two statements in the preceding paragraph requires a proof. In
particular, these statements are not obvious because of the potential discontinuity phenomenon that shows up in Lemma 
\ref{lem:psilem}.

Lei: The two statements have been removed. 
\color{black}
\fi

\subsection{Notation}

\label{subsec:notation}

%\black
We use upper-case letters, e.g., $X$, to denote a random variable
on 
%an alphabet 
%\color{magenta}
a finite alphabet
%\color{black}
$\mathcal{X}$. We use the lower-case letter $x$ to
denote a realization of $X$. We denote the distribution or the probability
mass function of $X$ as $P_{X}$, and use $Q_{X}$ to denote the
distribution of another r.v. on the same alphabet $\mathcal{X}$.
For brevity, the probability values $P_{X}(x)$ are sometimes written
as $P(x)$, when the subscript and the parameter are the same except
that the subscript is upper-case, and the parameter is lower-case.
We use $\mathbf{X}:=(X_{1},X_{2},...,X_{N})$ to denote a random vector.
We use the notation $A\leftrightarrow C\leftrightarrow B$ for a triple
of random variables $(A,B,C)$ to denote that $A$ and $B$ are conditionally
independent given $C$. We will also use notations $H_{Q}(X)$ or
$H(Q_{X})$ to denote the entropy of $X\sim Q_{X}$. If the distribution
is denoted by $P_{X}$, %sometimes we 
%\red
we sometimes %\black
write the entropy as $H(X)$ for brevity. %\red We will also occasionally use the shortened notation  $T$ to denote $T_{XY}$, viewed as a probability distribution on $\mathcal{X} \times \mathcal{Y}$.
We use $\supp(P_{X})$ to denote the support of $P_{X}$. %\black
The logarithm is taken to the natural base. %\blue 
Note that, as is the case for many other information-theoretic results,
the results in this paper can be viewed as independent of the choice
of the base of the logarithm as long as exponentiation is interpreted
as being with respect to the same base. 
%\color{magenta}
Also, for notational convenience, we will write $\mathrm{Unif}\left[1:e^{NR}\right]$ for a probability distribution that is uniform on 
$\left[\lceil e^{NR} \rceil\right]$,
where for a positive integer $n$
the notation $[n]$ denotes the set
$\{1, \ldots, n\}$.
%\color{black}

\iffalse
\color{red}
VA comment: I changed 
$\mathrm{Unif}\left[1:2^{NR}\right]$
to
$\mathrm{Unif}\left[1:e^{NR}\right]$
wherever it occurred.
\color{black}
\blue
(Lei: Your revision is fine. )
\black
\fi

%\color{magenta}
Since there are different notions of conditional R\'{e}nyi divergence in the literature,
we give a detailed description of the notion we use.
%\color{black}
Fix distributions $P_{X},Q_{X}$ on the same alphabet $\calX$. 
%\color{magenta}
For $s > 0$ the
%\color{black}
%The
{\em relative entropy} and the {\em R\'enyi divergence of order
$1+s$} are respectively defined as 
\begin{align}
D(P_{X}\|Q_{X}) & :=\sum_{x\in\mathrm{supp}(P_{X})}P_{X}(x)\log\frac{P_{X}(x)}{Q_{X}(x)}\label{eq:-19-1}\\*
D_{1+s}(P_{X}\|Q_{X}) & :=\frac{1}{s}\log\sum_{x\in\mathrm{supp}(P_{X})}P_{X}(x)^{1+s}Q_{X}(x)^{-s}.\label{eq:-40}
\end{align}
These are standard notions, see e.g. \cite{Erven}.
The conditional versions are respectively defined as 
\begin{align}
D(P_{Y|X}\|Q_{Y|X}|P_{X}) & :=D(P_{X}P_{Y|X}\|P_{X}Q_{Y|X})\\*
D_{1+s}(P_{Y|X}\|Q_{Y|X}|P_{X}) & :=D_{1+s}(P_{X}P_{Y|X}\|P_{X}Q_{Y|X}),
\end{align}
%\color{magenta}
the first of these being of course standard.
%\color{black}
%where the summations in \eqref{eq:-19-1} and \eqref{eq:-40} are
%taken over the elements in $\mathrm{supp}(P_{X})$.  
It is known
that $D_1(P_{X}\|Q_{X}):=\lim_{s\to0}D_{1+s}(P_{X}\|Q_{X})=D(P_{X}\|Q_{X})$ so a special
case of the R\'enyi divergence (or the conditional version) is the usual
relative entropy (respectively the conditional version).
%\color{magenta}
It can be checked that the data processing inequality for relative entropy extends to 
the R\'{e}nyi divergence, i.e. for $s\ge 0$,
\[
D_{1+s}(P_{XY}\|Q_{XY}) \ge D_{1+s}(P_{X}\|Q_{X}).
\]
%\color{black}

%\color{magenta}
The entropy of a random variable $X$ on a finite alphabet 
$\mathcal{X}$ with probability distribution $P_{X}$ can be written as
\[
H(X) := H(P_{X}) = \log |\mathcal{X}| - D(P_{X}\|U_{X}),
\]
where $|\mathcal{X}|$ denotes the cardinality of $\mathcal{X}$ and $U_{X}$ denotes the
uniform distribution on $\mathcal{X}$.
Thus for $s > 0$ the {\em R\'enyi entropy of order
$1+s$} is defined as
\[
H_{1+s}(X) := H_{1+s}(P_{X}) := \log |\mathcal{X}| - D_{1+s}(P_{X}\|U_{X}) = 
- \frac{1}{s}\log\sum_{x\in\mathrm{supp}(P_{X})}P_{X}(x)^{1+s}.
\]
This is a standard notion.
Note that if $X$ and $Y$ are independent then for all $s > 0$ we have
\[
H_{1+s}(XY) = H_{1+s}(X) + H_{1+s}(Y).
\]

For the conditional versions, for conditional entropy we have
\[
H(Y|X) = H(XY) - H(X) = H(P_{Y|X}| P_{X}) = \log |\mathcal{Y}| - D(P_{Y|X}\|U_{Y}|P_{X}).
\]
Thus for $s > 0$ the {\em conditional R\'enyi entropy of order
$1+s$} is defined as
\[
H_{1+s}(Y|X) := H_{1+s}(P_{Y|X}| P_{X}) := \log |\mathcal{Y}| - D_{1+s}(P_{Y|X}\|U_{Y}|P_{X}) = 
- \frac{1}{s}\log\sum_{x\in\mathrm{supp}(P_{X})}P_{X}(x) \sum_{y\in\mathrm{supp}(P_{Y})} P_{Y|X}(y|x)^{1+s}.
\]
It can be checked that we have 
$\lim_{s\to0}H_{1+s}(X) = H(X)$
and 
$\lim_{s\to0}H_{1+s}(Y|X)= H(Y|X)$.

With these definitions, as a caveat we note that while it is true that 
\[
H(Y|X) = \sum_{x\in\mathrm{supp}(P_{X})} P_{X}(x) H(Y|X = x),
\]
where $H(Y|X = x)$ denotes the entropy of the probability distribution $(P_{Y|X}(y|x), y \in \mathcal{Y})$,
for $s > 0$ we have in general that
\[
H_{1+s}(Y|X) \neq \sum_{x\in\mathrm{supp}(P_{X})} P_{X}(x) H_{1+s}(Y|X = x),
\]
where we will use the notation $H_{1+s}(Y|X = x)$ to denote the R\'{e}nyi entropy of 
the probability distribution $(P_{Y|X}(y|x), y \in \mathcal{Y})$ (similarly, for instance,
$H_{1+s}(Z|Y, X = x)$ will denote the conditional R\'{e}nyi entropy under the joint 
probability distribution $(P_{YZ|X}(y,z|x), (y,z) \in \mathcal{Y} \times \mathcal{Z})$).
Note that
$\lim_{s\to0}H_{1+s}(Y|X=x) = H(Y|X=x)$.

%\color{magenta}
Similarly the chain rule 
does not hold for
R\'{e}nyi entropy, i.e. for $s > 0$ 
in general we have
\[
H_{1+s}(YZ|X) \neq H_{1+s}(Y|X) + H_{1+s}(Z|XY).
\]
On the other hand, if $Z$ is independent of $(X,Y)$ then we have
\[
H_{1+s}(YZ|X) = H_{1+s}(Y|X) + H_{1+s}(Z).
\]
%\color{black}
\iffalse
\blue
Lei: The chain rule does not seem to hold for Renyi entropy. 
\black
\fi

In this document we do not need the R\'{e}nyi divergence and related notions for $s < 0$.
%\color{black}

\section{The Point-to-point Case}
 
For the sequential channel synthesis problem,
%, the following is one
%of our main results, in which 
%\color{magenta}
in this section
%\color{black}
we provide a single-letter characterization
of $\Gamma\left(\pi_{XY},P_{W}\right)$
in Theorem 
\ref{thm:sequentialCS},
%\color{magenta}
which is one of our main results.
%\color{black}
%\blue 
Define 
$\Psi:  \mathbb{R} \to [0,+\infty]$ 
%given 
by
% $\Psi:  \mathbb{R} \to [0,+\infty]$ given by\footnote{$\inf \emptyset := +\infty$.} 
\begin{equation}        \label{eq:psidef}
\Psi(t):=\min_{\substack{P_{UV},P_{B|XUV},P_{Y|BUV}:\\
H\left(U|V\right) \le  H\left(BU|XYV\right)+t
}
}D\left(P_{Y|XV}\|\pi_{Y|X}|\pi_{X}P_{V}\right),
\end{equation}
%\color{magenta}
where $B\in\mathcal{B}$, and all the entropies in \eqref{eq:psidef}
are evaluated at the joint distribution $\pi_{X}P_{UV}P_{B|XUV}P_{Y|BUV}$
and the distribution $P_{Y|XV}$ is also induced by this joint distribution.
%\color{black}
%\color{magenta}
Note that the minimum in \eqref{eq:psidef} is achieved
because a nonnegative lower semicontinuous function 
achieves its minimum on a compact set.
%Clearly $\Psi(t)$ is noincreasing in $t$.
%\color{black}
Denote $t_{\min}$ as the infimum of $t\in \mathbb{R}$ such that
$\Psi(t)<+\infty$. 
\begin{lem} \label{lem:psilem}
1) $\Psi(t)$ is convex and nonincreasing on $\mathbb{R}$. Moreover,  $\Psi(t)$ is equal to   $+\infty $ on $(-\infty,t_{\min})$, and   continuous on $(t_{\min},+\infty)$. \\
2) A sufficient condition for $t_{\min}<0$  is  the following assumption.

%\blue
Assumption 1: $|\mathcal{B}|\ge 2$ and there is at least one $y$ such that $\pi_{Y|X}(y|x)>0$ for all $x$ such that $\pi_{X}(x)>0$.  
\end{lem}
%\black

\begin{IEEEproof}
We first prove Statement 1). Since both the objective function and constraint functions are  linear in $P_V$ given $(P_{U|V},P_{B|XUV},P_{Y|BUV})$, $\Psi(t)$ is in fact a convex function. This can be shown by the standard argument that for two tuples   of r.v.'s $(X_1,V_1,U_1,B_1,Y_1)$ and $(X_2,V_2,U_2,B_2,Y_2)$, we can define a new r.v. $(X,U,B,Y):=(X_J,U_J,B_J,Y_J)$ and $V:=(V_J,J)$ where $J \sim \mathrm{Bern}(p)$ is independent of $V_1,V_2$). Then, the resultant objective function and constraint functions are the averages (with respect to $\mathrm{Bern}(p)$) of those for $(X_1,V_1,U_1,B_1,Y_1)$ and $(X_2,V_2,U_2,B_2,Y_2)$.
By the convexity, $\Psi(t)$ is continuous on $(t_{\min},+\infty)$. 

We next prove Statement 2). If we choose $U,V$ as constants, $B \sim \mathrm{Unif}(\mathcal{B}),X$ are mutually independent, and  $Y=y$ as constant as well (here $y$ is the element given in the lemma), then this set of distributions is feasible if $t> -\log |\mathcal{B}|$ and the resultant value is finite. Hence, $t_{\min}<0$. 
\end{IEEEproof}
Denote $\Delta\left(\pi_{XY},P_{W}\right):=\Psi(H(W))$.
% \color{magenta}
% Also let
% \[
% \Delta^{+}\left(\pi_{XY},P_{W}\right):=
% \inf_{\substack{P_{UV},P_{B|XUV},P_{Y|BUV}:\\
% H\left(U|V\right) <  H\left(BU|XYV\right)+ H(W)
% }
% }D\left(P_{Y|XV}\|\pi_{Y|X}|\pi_{X}P_{V}\right), 
% \]
% where $B\in\mathcal{B}$, and all the entropies 
% are evaluated at the joint distribution $\pi_{X}P_{UV}P_{B|XUV}P_{Y|BUV}$
% and the distribution $P_{Y|XV}$ is also induced by this joint distribution.
% From Lemma \ref{lem:psilem} we have 
% $\Delta^{+}\left(\pi_{XY},P_{W}\right) \ge 
% \Delta\left(\pi_{XY},P_{W}\right)$ with equality except possibly when $H(W) = t_{\min}$.
% \color{black} 
The proof of the following theorem is provided in
Appendix \ref{sec:Proof-of-Theorem}. 
\begin{thm}
\label{thm:sequentialCS} %Given $\mathcal{B}$, 
%\blue 
Under Assumption 1, we have 
%\color{magenta} 
\begin{equation} 
\Gamma\left(\pi_{XY},P_{W}\right) = \Delta\left(\pi_{XY},P_{W}\right).\label{eq:-18-1}
\end{equation}
% \begin{equation}
% \Delta\left(\pi_{XY},P_{W}\right) \le
% \Gamma\left(\pi_{XY},P_{W}\right) \le \Delta^{+}\left(\pi_{XY},P_{W}\right).\label{eq:-18-1}
% \end{equation}
%\color{black}
Furthermore, it suffices to restrict the cardinality of $\mathcal{U}$
and $\mathcal{V}$ 
%\color{magenta}
in the calculation of
$\Delta\left(\pi_{XY},P_{W}\right)$
%\color{black}
such that 
%\blue 
$\left|\mathcal{V}\right|\leq 2$
and $\left|\mathcal{U}\right|\leq 2\left|\mathcal{X}\right|\left|\mathcal{Y}\right|$. 
%\black
% \color{magenta}
% Note that if
% \footnote{The condition $H(W)\neq t_{\min}$   holds trivially  if  $t_{\min}<0$.}  $H(W)\neq t_{\min}$, 
% and in particular if
% there at least one $y$ such that $\pi_{Y|X}(y|x)>0$ for all $x$ such that $\pi_{X}(x)>0$,
% then
% \eqref{eq:-18-1}
% completely characterizes
% $\Gamma\left(\pi_{XY},P_{W}\right)$.
% \color{black}
\end{thm}
\iffalse
\blue 
Lei: The cardinality bound was improved, and the proof is given in Appendix A. 
\black
\fi
\iffalse
\color{red}
VA comment: The claimed cardinality bounds on the auxiliaries do not seem to have been proved in the appendix.
\color{black}
\fi
\begin{rem}
Note that $\Delta\left(\pi_{XY},P_{W}\right)$ depends on $P_{W}$
only through its entropy $H\left(W\right)$. 
\end{rem}
%If $P_{W}$ is degenerate, and at the same time, 
We next consider the case in which the stochastic encoder $P_{B_{k}|W^{k}X^{k}B^{k-1}}$
and decoder $P_{Y_{k}|W^{k}B^{k}Y^{k-1}}$ are respectively replaced
by $P_{B_{k}|X^{k}}$ and $P_{Y_{k}|B^{k}Y^{k-1}}$. In other words,
in this case, it is not allowed to extract common randomness for the
communication at the $k$-th epoch from the previous communication
bits $B^{k-1}$
%\color{magenta}
and there is no externally provided common randomness.
%\color{black}
We next show that a symbol-by-symbol mapping suffices
to achieve the optimal KL divergence for this case, which we denote
by $\widetilde{\Gamma}_{0}\left(\pi_{XY}\right)$, as shown in the
following result. 
\begin{rem}
Note that $\widetilde{\Gamma}_{0}\left(\pi_{XY}\right)$ is a priori
smaller than $\Gamma_{0}\left(\pi_{XY}\right)$, because the latter
allows for stochastic encoders of the form $P_{B_{k}|X^{k}B^{k-1}}$
which, with decoders of the form $P_{Y_{k}|B^{k}Y^{k-1}}$, allows
for the possibility of extracting common randomness from the communication. 
\end{rem}
\begin{thm}
\label{thm:where--is}If the stochastic encoder $P_{B_{k}|W^{k}X^{k}B^{k-1}}$
and decoder $P_{Y_{k}|W^{k}B^{k}Y^{k-1}}$ are respectively replaced
by $P_{B_{k}|X^{k}}$ and $P_{Y_{k}|B^{k}Y^{k-1}}$, then 
\[
\widetilde{\Gamma}_{0}\left(\pi_{XY}\right)=\widetilde{\Delta}\left(\pi_{XY}\right):=\inf_{P_{B|X},P_{Y|B}}D\left(P_{Y|X}\|\pi_{Y|X}|\pi_{X}\right)
\]
where $B\in\mathcal{B}$ and $P_{Y|X}$ is induced by the joint distribution
$\pi_{X}P_{B|X}P_{Y|B}$. 
\end{thm}
% \cor
% Note to Lei: Removed the phrase ``If $P_{W}$ is degenerate" from the preceding remark since the theorem works even if $\{W_i\}$ is provided, as long as Alice and Bob do not use it.
% \cob

\begin{IEEEproof}
It is easy to see that 
%$\widetilde{\Gamma}\left(\pi_{XY}\right)\leq\widetilde{\Delta}\left(\pi_{XY}\right)$
%\color{magenta}
$\widetilde{\Gamma}_{0}\left(\pi_{XY}\right)\leq\widetilde{\Delta}\left(\pi_{XY}\right)$
%\color{black}
since $\widetilde{\Delta}\left(\pi_{XY}\right)$ is achievable by
a communication scheme consisting of symbol-by-symbol mappings.

On the other hand, 
\begin{align}
 & D\left(P_{Y^{n}|X^{n}}\|\pi_{Y|X}^{n}|\pi_{X}^{n}\right)\nonumber \\
 & =D\left(\prod_{k=1}^{n}P_{Y_{k}|X^{k}Y^{k-1}}\|\pi_{Y|X}^{n}|\pi_{X}^{n}\right)\nonumber \\
 & =\sum_{k=1}^{n}D\left(P_{Y_{k}|X^{k}Y^{k-1}}\|\pi_{Y|X}|\pi_{X}^{k}P_{Y^{k-1}|X^{k}}\right)\nonumber \\
 & \geq\sum_{k=1}^{n}D\left(P_{Y_{k}|X^{k}}\|\pi_{Y|X}|\pi_{X}^{k}\right)\label{eq:-4}\\
 & \geq\sum_{k=1}^{n}\min_{x^{k-1}}D\left(P_{Y_{k}|X_{k},X^{k-1}=x^{k-1}}\|\pi_{Y|X}|\pi_{X}^{k}\right)\label{eq:-5}
\end{align}
where \eqref{eq:-4} follows %since the KL divergence between two joint
%distributions is not smaller than that between two marginal distributions,
from the convexity of $D(p\|q)$ in the pair $(p,q)$, and \eqref{eq:-5}
follows since 
\begin{align*}
D\left(P_{Y_{k}|X^{k}}\|\pi_{Y_{k}|X^{k}}|\pi_{X^{k}}\right) & =\mathbb{E}_{X^{k-1}\sim\pi_{X}^{k-1}}D\left(P_{Y_{k}|X_{k},X^{k-1}=x^{k-1}}\|\pi_{Y|X}|\pi_{X}\right)\\
 & \geq\min_{x^{k-1}}D\left(P_{Y_{k}|X_{k},X^{k-1}=x^{k-1}}\|\pi_{Y|X}|\pi_{X}\right).
\end{align*}
It is easy to verify that (note that the following does not hold if
we consider encoder $P_{B_{k}|X^{k}B^{k-1}}$) 
\begin{align}
P_{X^{k}B_{k}Y_{k}}\left(x^{k},b_{k},y_{k}\right) & =\sum_{b^{k-1},y^{k-1}}\pi_{X}^{k}\left(x^{k}\right)P_{B^{k-1}|X^{k-1}}\left(b^{k-1}|x^{k-1}\right)P_{Y^{k-1}|B^{k-1}X^{k-1}}(y^{k-1}|b^{k-1},x^{k-1}) \nonumber\\
 & \qquad\times P_{B_{k}|X^{k}}\left(b_{k}|x^{k}\right)P_{Y_{k}|B^{k}Y^{k-1}}\left(y_{k}|b^{k},y^{k-1}\right) \label{eq:factor}\\
 & =\pi_{X}^{k}\left(x^{k}\right)P_{B_{k}|X^{k}}\left(b_{k}|x^{k}\right)P_{Y_{k}|B_{k}X^{k-1}}\left(y_{k}|b_{k},x^{k-1}\right).
\end{align}
\iffalse
\color{red}
VA question: I am not sure I understand why the preceding claim is true. Note that
\[
P_{X^{k}B_{k}Y_{k}}\left(x^{k},b_{k},y_{k}\right)
= \pi_{X}^{k}\left(x^{k}\right)P_{B_{k}|X^{k}}\left(b_{k}|x^{k}\right)P_{Y_{k}|B_{k}X^{k}}\left(y_{k}|b_{k},x^{k}\right),
\]
so what is being claimed is basically
\[
P_{Y_{k}|B_{k}X^{k-1}}\left(y_{k}|b_{k},x^{k-1}\right)
=
P_{Y_{k}|B_{k}X^{k}}\left(y_{k}|b_{k},x^{k}\right),
\]
i.e. the Markov chain
\[
X_k \leftrightarrow \left(B_{k},X^{k-1}\right)
\leftrightarrow Y_{k}.
\]
But it is not clear intuitively why this should be true. Since $Y_{k}$ is chosen based on 
$\left(B^{k},Y^{k-1}\right)$ what matters is 
whether the Markov chain
$\left(B^{k},Y^{k-1}\right) \leftrightarrow
\left(B_{k},X^{k-1}\right)
\leftrightarrow
X_{k}$
holds and it seems to me that this need not be the
case because $B_{k}$ is allowed to depend on all of $X^{k}$.

(Lei: The Markov chains you mentioned follow from the factorization given in \eqref{eq:factor}, i.e., 
\[
P_{X^k B^k Y^k}= \pi_{X}^{k} P_{B^{k-1}|X^{k-1}} P_{Y^{k-1}|B^{k-1}X^{k-1}}  P_{B_{k}|X^{k}} P_{Y_{k}|B^{k}Y^{k-1}}.
 \]
 Taking the conditional distribution, we have 
 \[
P_{Y^k B^{k-1} | X^k B_k }= P_{B^{k-1}|X^{k-1}} P_{Y^{k-1}|B^{k-1}X^{k-1}}  P_{Y_{k}|B^{k}Y^{k-1}}.
 \]
 Note that there is no $X_k$ appearing at the RHS above, which 
 implies $Y^k B^{k-1} \leftrightarrow X^{k-1} B_k \leftrightarrow X_k$.)
\color{black}
\fi
Let $\hat{x}^{k-1}$ be the optimal sequence that attains the minimum
in \eqref{eq:-5}. Then given $X^{k-1}=\hat{x}^{k-1}$, 
\begin{align*}
P_{X_{k}B_{k}Y_{k}|X^{k-1}}\left(x_{k},b_{k},y_{k}|\hat{x}^{k-1}\right) & =\pi_{X}\left(x_{k}\right)P_{B_{k}|X^{k}}\left(b_{k}|x_{k},\hat{x}^{k-1}\right)P_{Y_{k}|B_{k}X^{k-1}}\left(y_{k}|b_{k},\hat{x}^{k-1}\right).
\end{align*}
By identifying $P_{B|X}=P_{B_{k}|X_{k},X^{k-1}=\hat{x}^{k-1}},P_{Y|B}=P_{Y_{k}|B_{k},X^{k-1}=\hat{x}^{k-1}}$,
we have 
%$\widetilde{\Gamma}\left(\pi_{XY}\right)\geq\widetilde{\Delta}\left(\pi_{XY}\right)$.
%\color{magenta}
$\widetilde{\Gamma}_{0}\left(\pi_{XY}\right)\geq\widetilde{\Delta}\left(\pi_{XY}\right)$.
%\color{black}
\end{IEEEproof}

\section{The Broadcast Case}

We now consider the sequential channel synthesis problem over a noiseless
broadcast channel. Let $\mathcal{W}$, $\hat{\mathcal{W}}$, $\mathcal{X}$,
$\mathcal{Y}$, $\mathcal{Z}$ and $\mathcal{B}$ be finite sets.
%\color{magenta}
Assume that $|\mathcal{B}| \ge 2$.
%\color{black}
Assume that a sender Alice and two receivers Bob and Charles share
a common random sequence $W^{k}$; in addition to this, Alice and
Bob also share another common random sequence $\hat{W}^{k}$. Here
$\{W_{i}\}$ is an i.i.d. sequence of random variables taking values
in $\mathcal{W}$ with each $W_{i}\sim P_{W}$ and $\{\hat{W}_{i}\}$
is an i.i.d. sequence of random variables taking values in $\hat{\mathcal{W}}$
with each $\hat{W}_{i}\sim P_{\hat{W}}$. %and $\{W_i\}$ and $\{\hat{W}_i\}$ are independent. 
There is also a sequence of random variables $\{X_{i}\}$ taking values
in $\mathcal{X}$, with $X_{i}\sim P_{X}$. We assume that $\{X_{i}\}$,
$\{W_{i}\}$ and $\{\hat{W}_{i}\}$ are mutually independent. The
sequence $\{X_{i}\}$ is called the source sequence
%\color{magenta}
and is observed only by Alice.
%\color{black}

At the epoch $k$, upon observing the random sequences $\left(W^{k},\hat{W}^{k}\right)$,
the source sequence $X^{k}$, and previous communication random variables
$B^{k-1}$, Alice generates $B_{k}\in\mathcal{B}$ by using a random
mapping with conditional distribution $P_{B_{k}|W^{k}\hat{W}^{k}X^{k}B^{k-1}}$,
and then sends $B_{k}$ to Bob and Charles. Upon observing $W^{k},\hat{W}^{k}$,
$B^{k}$, and previous outputs $Y^{k-1}$, Bob generates $Y_{k}$
by using a random mapping with conditional distribution $P_{Y_{k}|W^{k}\hat{W}^{k}B^{k}Y^{k-1}}$.
Upon observing $W^{k}$, $B^{k}$, and the previous outputs $Z^{k-1}$,
Charles generates $Z_{k}$ by using a random mapping with conditional
distribution $P_{Z_{k}|W^{k}B^{k}Z^{k-1}}$. Given a target broadcast
channel $\pi_{YZ|X}$, the goal is for Alice, Bob, and Charles to
cooperate in this sequential manner to minimize the KL divergence
$D\left(P_{Y^{n}Z^{n}|X^{n}}\|\pi_{YZ|X}^{n}|\pi_{X}^{n}\right)$
between the synthesized joint distribution $\pi_{X}^{n}P_{Y^{n}Z^{n}|X^{n}}$
and the target joint distribution $\pi_{X}^{n}\pi_{YZ|X}^{n}$. Here
the broadcast channel from Alice to Bob and Charles that has been
synthesized is 
\begin{align*}
P_{Y^{n}Z^{n}|X^{n}}\left(y^{n},z^{n}|x^{n}\right) & :=\sum_{w^{n},\hat{w}^{n},b^{n}}P_{W}^{n}\left(w^{n}\right)P_{\hat{W}}^{n}\left(\hat{w}^{n}\right)\prod_{k=1}^{n}P_{B_{k}|W^{k}\hat{W}^{k}X^{k}B^{k-1}}\left(b_{k}|w^{k},\hat{w}^{k},x^{k},b^{k-1}\right)\\
 & \qquad\times\prod_{k=1}^{n}P_{Y_{k}|W^{k}\hat{W}^{k}B^{k}Y^{k-1}}\left(y_{k}|w^{k},\hat{w}^{k},b^{k},y^{k-1}\right)\prod_{k=1}^{n}P_{Z_{k}|W^{k}B^{k}Z^{k-1}}\left(z_{k}|w^{k},b^{k},z^{k-1}\right).
\end{align*}
We are interested in characterizing 
\begin{equation}
\Gamma\left(\pi_{XYZ},P_{W}P_{\hat{W}}\right):=\lim_{n\to\infty}\inf_{\left\{ \left(P_{B_{k}|W^{k}\hat{W}^{k}X^{k}B^{k-1}},P_{Y_{k}|W^{k}\hat{W}^{k}B^{k}Y^{k-1}},P_{Z_{k}|W^{k}B^{k}Z^{k-1}}\right)\right\} _{k=1}^{n}}\frac{1}{n}D\left(P_{Y^{n}Z^{n}|X^{n}}\|\pi_{YZ|X}^{n}|\pi_{X}^{n}\right).\label{eq:-3-1}
\end{equation}

For this sequential broadcast channel synthesis problem, we prove
the following result. The proof is provided in Appendix \ref{sec:Proof-of-Theorem-broadcast}. 
\begin{thm}
%\blue 
Assume 
$|\mathcal{B}|\ge 2$ and there is at least one  pair $(y,z)$ such that $\pi_{YZ|X}(y,z|x)>0$ for all $x$ such that $\pi_{X}(x)>0$. 
%\black
%\color{magenta}
Then we have
%\color{black}
\label{thm:sequentialCS-broadcast-1} 
\begin{equation}
\Delta\left(\pi_{XYZ},P_{W}P_{\hat{W}}\right)\leq\Gamma\left(\pi_{XYZ},P_{W}P_{\hat{W}}\right)\leq\hat{\Delta}\left(\pi_{XYZ},P_{W}P_{\hat{W}}\right)\label{eq:-18-1-4-2}
\end{equation}
where 
\begin{equation}
\Delta\left(\pi_{XYZ},P_{W}P_{\hat{W}}\right):=\min_{\substack{P_{U\hat{U}V},P_{B|XU\hat{U}V},P_{Y|BU\hat{U}V},P_{Z|BUV}:\\
H\left(U|V\right)\le H(W)+H\left(BU|XYZV\right),\\
H\left(U\hat{U}|V\right)\le H(\hat{W})+H(W)+H\left(BU\hat{U}|XYZV\right)
}
}D\left(P_{YZ|XV}\|\pi_{YZ|X}|\pi_{X}P_{V}\right),\label{eq:-18-1-4-2-1}
\end{equation}
and $\hat{\Delta}\left(\pi_{XYZ},P_{W}P_{\hat{W}}\right)$ is defined
as the expression identical to $\Delta\left(\pi_{XYZ},P_{W}P_{\hat{W}}\right)$
except that $I\left(B;\hat{U}|XYZUV\right)$ is additionally added
to the LHS in the first constraint. 
%and 
% both ``$\le$'' in the constraints are replaced by ``$<$'',
% \color{magenta}
% and the ``min" is replaced by ``inf".
% \color{black}
Here all the entropies are evaluated
at the joint distribution \[
\pi_{X}P_{U\hat{U}V}P_{B|XU\hat{U}V}P_{Y|BU\hat{U}V}P_{Z|BUV}
\]
and the distribution $P_{YZ|XV}$ is also induced by this joint distribution. 
%\blue 
Furthermore, it suffices to restrict the cardinality of
$\mathcal{V}, \mathcal{U}$
and $\hat{\mathcal{U}}$
%\color{magenta}
in the calculation of
$\Delta\left(\pi_{XYZ},P_{W}P_{\hat{W}}\right)$
%\color{black}
such that $\left|\mathcal{V}\right|\leq 3$,  $\left|\mathcal{U}\right|\leq 3(|\mathcal{X}||\mathcal{Y}||\mathcal{Z}|+1)$,  
and $|\hat{\mathcal{U}}|\leq 3(|\mathcal{X}||\mathcal{Y}||\mathcal{Z}|+1)|\mathcal{B}||\mathcal{X}||\mathcal{Y}||\mathcal{Z}|$. Similarly, it suffices to restrict the cardinality of
$\mathcal{V}, \mathcal{U}$
and $\hat{\mathcal{U}}$
%\color{magenta}
in the calculation of
$\hat{\Delta}\left(\pi_{XYZ},P_{W}P_{\hat{W}}\right)$
%\color{black}
such that $\left|\mathcal{V}\right|\leq 3$,  $\left|\mathcal{U}\right|\leq 3(|\mathcal{X}||\mathcal{Y}||\mathcal{Z}|+1)$,  
and $|\hat{\mathcal{U}}|\leq 3(|\mathcal{X}||\mathcal{Y}||\mathcal{Z}|+1)(|\mathcal{B}||\mathcal{X}||\mathcal{Y}||\mathcal{Z}|+1)$. 
\end{thm}
%\black
\iffalse
\color{red}
VA question: Cardinality bounds on the auxiliaries seems to be missing in the statement of the theorem as
opposed to the other theorems.
\color{black}

\blue 
Lei: I have added the cardinality bounds and the corresponding proofs. 
\black 
\fi
\section{The Interactive Communication Case}

We now consider the sequential channel synthesis problem over a noiseless
\emph{two-way} channel. Let $\{(S_{k},X_{k})\}$ be a memoryless source
with $(S_{k},X_{k})\sim\pi_{SX}$ for all $k$. At epoch $k$, upon
observing the common random sequence $W^{k}$, the source sequence
$S^{k}$,  previous communication random variables $\left(A^{k-1},B^{k-1}\right)$, and the previous output $Y^{k-1}$,
Alice generates 
%one random r.v. 
$A_{k}\in\mathcal{A}$ by using a
random mapping $P_{A_{k}|S^{k}A^{k-1}B^{k-1}Y^{k-1}W^{k}}$, and then sends
it to Bob. At the same epoch, upon observing the common random sequence
$W^{k}$, the source sequence $X^{k}$,  previous communication
random variables $\left(A^{k-1},B^{k-1}\right)$, , and the previous output $Z^{k-1}$, Bob generates 
%one random r.v. 
$B_{k}\in\mathcal{B}$ by using a random mapping $P_{B_{k}|X^{k}A^{k-1}B^{k-1}Z^{k-1}W^{k}}$,
and then sends it to Alice.

\iffalse
\color{red}
VA question: Since Alice has access to $Y^{k-1}$ at time $k$ and Bob has access to 
$Z^{k-1}$ at time $k$ (see the next paragraph) it seems strange that $A_{k}$ 
is generated via $P_{A_{k}|S^{k}A^{k-1}B^{k-1}W^{k}}$ instead of 
$P_{A_{k}|S^{k}Y^{k-1}A^{k-1}B^{k-1}W^{k}}$ and similarly that 
$B_{k}$ is generated via $P_{B_{k}|X^{k}A^{k-1}B^{k-1}W^{k}}$
instead of $P_{B_{k}|X^{k}Z^{k-1}A^{k-1}B^{k-1}W^{k}}$. It seems that if one generalizes
the formulation in this way the answers may change. Maybe some comment should be made about this.
\color{black}
\blue
(Lei: Your suggestion is great. I have revised $P_{A_{k}|S^{k}A^{k-1}B^{k-1}W^{k}}$ to 
$P_{A_{k}|S^{k}Y^{k-1}A^{k-1}B^{k-1}W^{k}}$, and also revised the corresponding parts in the proof. However, the result in  Theorem 4
remains unchanged.)
\black
\fi
Also at epoch $k$, upon observing $W^{k}$, $A^{k},B^{k}$, source
sequence $S^{k}$, and previous outputs $Y^{k-1}$, Alice generates
%a r.v. 
$Y_{k}$ by using a random mapping $P_{Y_{k}|A^{k}B^{k}S^{k}Y^{k-1}W^{k}}$.
Upon observing $W^{k}$, $A^{k},B^{k}$, source sequence $X^{k}$,
and previous outputs $Z^{k-1}$, Bob generates a r.v. $Z_{k}$ by
using a random mapping $P_{Z_{k}|A^{k}B^{k}X^{k}Z^{k-1}W^{k}}$. Given
a target channel $\pi_{YZ|SX}$, 
%Alice, Bob, and Charlie 
%\color{magenta}
Alice and Bob
%\color{black}
cooperate
to minimize the KL divergence $D\left(P_{Y^{n}Z^{n}|S^{n}X^{n}}\|\pi_{YZ|SX}^{n}|\pi_{SX}^{n}\right)$
between the synthesized channel and the target channel. We are interested
in characterizing 
\begin{equation}
\Gamma\left(\pi_{SXYZ},P_{W}\right):=\lim_{n\to\infty}\inf_{\left\{ \left(\substack{P_{A_{k}|S^{k}A^{k-1}B^{k-1}Y^{k-1}W^{k}},P_{B_{k}|X^{k}A^{k-1}B^{k-1}Z^{k-1}W^{k}},\\
P_{Y_{k}|A^{k}B^{k}S^{k}Y^{k-1}W^{k}},P_{Z_{k}|A^{k}B^{k}X^{k}Z^{k-1}W^{k}}
}
\right)\right\} _{k=1}^{n}}\frac{1}{n}D\left(P_{Y^{n}Z^{n}|S^{n}X^{n}}\|\pi_{YZ|SX}^{n}|\pi_{SX}^{n}\right).\label{eq:-3-1-5}
\end{equation}

%\color{magenta}
Let
\[
\Delta\left(\pi_{SXYZ},P_{W}\right):=\min_{\substack{P_{UV},P_{A|SUV},P_{B|XUV},P_{Y|ABUV},P_{Z|ABUV}:\\
H\left(U|V\right)\leq H\left(ABU|SXYZV\right)+H\left(W\right)
}
}D\left(P_{YZ|SXV}\|\pi_{YZ|SX}|\pi_{SX}P_{V}\right),
\]
% and
% \[
% \Delta^{+}\left(\pi_{SXYZ},P_{W}\right):=\inf_{\substack{P_{UV},P_{A|SUV},P_{B|XUV},P_{Y|ABUV},P_{Z|ABUV}:\\
% H\left(U|V\right) < H\left(ABU|SXYZV\right)+H\left(W\right)
% }
% }D\left(P_{YZ|SXV}\|\pi_{YZ|SX}|\pi_{SX}P_{V}\right),
% \]
where $A\in\mathcal{A},B\in\mathcal{B}$ and all the entropies above
are evaluated at the joint distribution $\pi_{SX}P_{UV}P_{B|XUV}P_{A|SUV}P_{Y|ABUV}P_{Z|ABUV}$
and the distribution $P_{YZ|SXV}$ is also induced by this joint distribution.
Note that these expressions depend on $P_{W}$ only through $H(W)$.
% Then 
% we have 
% $\Delta^{+}\left(\pi_{SXYZ},P_{W}\right) \ge
% \Delta\left(\pi_{SXYZ},P_{W}\right)$ and,
% based on an analog of 
% Lemma \ref{lem:psilem},
% we have equality except possibly for one value of $H(W)$.
% Further, if
% there at least one $(y,z)$ such that $\pi_{YZ|SX}(y,z|s,x)>0$ for all $(s,x)$ such that $\pi_{SX}(s,x)>0$ then 
% we have equality of 
% $\Delta^{+}\left(\pi_{SXYZ},P_{W}\right)$
% and
% $\Delta\left(\pi_{SXYZ},P_{W}\right)$.
%\color{black}

For this interactive version of sequential channel synthesis problem,
we prove the following result. The proof is provided in Appendix \ref{sec:Proof-of-Theorem-interaction}. 
\begin{thm}
\label{thm:sequentialCS-interaction} 
%\begin{equation}
%\Gamma\left(\pi_{SXYZ},P_{W}\right)=\Delta\left(\pi_{SXYZ},P_{W}\right):=\min_{\substack{P_{UV},P_{A|SUV},P_{B|XUV},P_{Y|ABUV},P_{Z|ABUV}:\\
%H\left(U|V\right)\leq H\left(ABU|SXYZV\right)+H\left(W\right)
%}
%}D\left(P_{YZ|SXV}\|\pi_{YZ|SX}|\pi_{SX}P_{V}\right)\label{eq:-18-1-4-1}
%\end{equation}
%\color{blue}
Assume that $|\mathcal{A}|, |\mathcal{B}|\ge 2$ and there is at least one $(y,z)$ such that $\pi_{YZ|SX}(y,z|s,x)>0$ for all $(s,x)$ such that $\pi_{SX}(s,x)>0$.   We have
% \begin{equation}
% \Delta\left(\pi_{SXYZ},P_{W}\right)
% \le 
% \Gamma\left(\pi_{SXYZ},P_{W}\right) \le \Delta^{+} \left(\pi_{SXYZ},P_{W}\right).
% \label{eq:-18-1-4-1}
% \end{equation}
\begin{equation}
\Gamma\left(\pi_{SXYZ},P_{W}\right) = \Delta \left(\pi_{SXYZ},P_{W}\right).
\label{eq:-18-1-4-1}
\end{equation}
%color{black}
%where $A\in\mathcal{A},B\in\mathcal{B}$ and all the entropies above
%are evaluated at the joint distribution $\pi_{SX}P_{UV}P_{B|XUV}P_{A|SUV}P_{Y|ABUV}P_{Z|ABUV}$
%and the distribution $P_{YZ|SXV}$ is also induced by this joint distribution.
Furthermore, 
in the calculation of 
$\Delta\left(\pi_{SXYZ},P_{W}\right)$
\color{black}
it suffices to restrict the cardinality of $\mathcal{U}$
and $\mathcal{V}$ such that
%\blue 
$\left|\mathcal{V}\right|\leq 2$
and $\left|\mathcal{U}\right|\leq 2 \left|\mathcal{S}\right|\left|\mathcal{X}\right|\left|\mathcal{Y}\right|\left|\mathcal{Z}\right|$.
%\black 
% \color{magenta}
% The bounds in
% \eqref{eq:-18-1-4-1} completely 
% characterize
% $\Gamma\left(\pi_{SXYZ},P_{W}\right)$ except possibly for one value of $H(W)$ and 
% in particular completely characterize
% $\Gamma\left(\pi_{SXYZ},P_{W}\right)$ if
% there is at least one $(y,z)$ such that $\pi_{YZ|SX}(y,z|s,x)>0$ for all $(s,x)$ such that $\pi_{SX}(s,x)>0$. 
\color{black}
\end{thm}

\appendices{}

\section{\label{sec:Proof-of-Theorem}Proof of Theorem \ref{thm:sequentialCS}}
\subsection{Cardinality Bounds}
%\blue 
To prove the claimed cardinality bounds  for $\Delta\left(\pi_{XY},P_{W}\right)$, it suffices to prove that the same 
cardinality bounds hold for $\Psi(t)$ with $t\ge 0$.
Note that the constraint in \eqref{eq:psidef} can be rewritten as 
$H(XY|V)-H(BXY|UV) \le t$. By the support lemma in \cite[Appendix C]{Gamal}, the cardinality of $\mathcal{V}$ can be upper bounded by 
$2$, without changing the constraint function and the objective function (both of which are linear in $P_V$). 
% The joint distribution can be still factorized as   $\pi_{X}P_{UV}P_{B|XUV}P_{Y|BUV}$.

Applying the support lemma in \cite[Appendix C]{Gamal} again, for each $v$, we can restrict the size of the  support  of $P_{U|V=v}$ no larger  than $|\mathcal{X}||\mathcal{Y}|$ 
without changing the linear functionals  $P_{XY|V=v}$ and $H(BXY|U,V=v)$, and hence also without changing the constraint function and the objective function. 
Therefore,  the cardinality of $\mathcal{U}$ can be upper bounded by
$2|\mathcal{X}||\mathcal{Y}|$.  
%\black
\subsection{Achievability}

To prove the achievability part, i.e., $\Gamma\left(\pi_{XY},P_{W}\right)\leq\Delta\left(\pi_{XY},P_{W}\right)$,
% Observe that the r.v. $V$ in the definition of $\Delta\left(\pi_{XY},P_{W}\right)$
% is a time-sharing r.v., which means that to prove $\Gamma\left(\pi_{XY},P_{W}\right)\leq\Delta\left(\pi_{XY},P_{W}\right)$,
we first  prove 
\begin{equation}
\Gamma\left(\pi_{XY},P_{W}\right)\leq\overline{\Delta}\left(\pi_{XY},P_{W}\right):=\inf_{\substack{P_{U},P_{B|XU},P_{Y|BU}:\\
H\left(U\right) < H\left(BU|XY\right)+H\left(W\right)
}
}D\left(P_{Y|X}\|\pi_{Y|X}|\pi_{X}\right).\label{eq:-18-1-1}
\end{equation}
\iffalse
\color{red}
VA question:
Suppose that in the expression on the right hand side of \eqref{eq:-18-1-1} I take $U$ to be a constant
(trivial) and $B = Y$. Then the condition 
$H\left(U\right) < H\left(BU|XY\right)+H\left(W\right)$ is clearly satisfied if $H(W) > 0$. But in this case 
we have 
$D\left(P_{Y|X}\|\pi_{Y|X}|\pi_{X}\right) = 0$. This seems strange.

(Lei: The explanation in your latest email is correct.) 
\color{black}
\fi
\iffalse
\color{red}
VA question:
I am not sure this is obvious. The condition
\[
H\left(U|V\right)\leq H\left(BU|XYV\right)+H\left(W\right)
\]
does not imply that for every $v$ we have
\[
H\left(U|V = v\right)\leq H\left(BU|XY,V = v\right)+H\left(W\right).
\]
Therefore focusing on $\overline{\Delta}\left(\pi_{XY},P_{W}\right)$
may not be enough to control
$\Delta\left(\pi_{XY},P_{W}\right)$. If the claim
is true more details may be needed.
(Lei: You are right. These two conditions are not equivalent. Thanks for pointing out this obvious error. It has been fixed by the argument given at the end of the achievability proof.)
\color{black}
\fi
% We next prove $\Gamma\left(\pi_{XY},P_{W}\right)\leq\overline{\Delta}\left(\pi_{XY},P_{W}\right)$.

Let $\left(Q_{U},Q_{B|XU},Q_{Y|BU}\right)$ be a tuple
%of optimal distributions that attains the infimum in 
%\color{magenta}
that satisfy the constraints in the expression on the RHS of
%\color{black}
\eqref{eq:-18-1-1}. 
%(If the infimum is not attained, then we can assume this tuple approximately attains the infimum within $\epsilon$ gap, and finally, let  $\epsilon\to 0$.)  
%Here
%\color{magenta}
For the achievability proof
%\color{black}
we will adopt block-by-block codes. For brevity, for a sequence of
r.v.'s $\left\{ Z_{i}\right\} $, we denote $\mathbf{Z}:=Z^{N}$ and
$\mathbf{Z}_{k}:=Z_{\left(k-1\right)N+1}^{kN}$. 
%\color{magenta}
We will also use the notation $Z_{ki}$ for $Z_{(k-1)N+ i}$ for $k \ge 1$ and $1 \le i \le N$ when
$N$ is known from the context.
%\color{black}
Let $\mathcal{C}:=\left\{ \mathbf{M}\left(\mathbf{b},\mathbf{w}\right):\left(\mathbf{b},\mathbf{w}\right)\in\mathcal{B}^{N}\times\mathcal{W}^{N}\right\} $
be a 
%common 
%\color{magenta}
%(between the sender and the receiver)
%\color{black}
random binning codebook where $\mathbf{M}\left(\mathbf{b},\mathbf{w}\right)\sim\mathrm{Unif}\left[1:e^{NR}\right]$
are generated independently. Let $\mathcal{C}_{k},k=1,2,...$ be independent
copies of $\mathcal{C}$. The codebook $\mathcal{C}_{k}$ will be
used to generate a nearly uniform r.v. from the previous block of
communication bits $\mathbf{B}_{k-1}$ and the common randomness $\mathbf{W}_{k-1}$.
Let $\hat{\mathcal{C}}:=\left\{ \mathbf{U}\left(m\right):m\in\left[1:e^{NR}\right]\right\} $
be another 
%common 
random codebook, where $\mathbf{U}\left(i\right)\sim\widetilde{Q}_{\mathbf{U}}$
are generated independently with $\widetilde{Q}_{\mathbf{U}}$ denoting
the following truncated product distribution: 
\[
\widetilde{Q}_{\mathbf{U}}=\frac{Q_{U}^{N}1_{\mathcal{T}_{\epsilon}^{(N)}\left(Q_{U}\right)}}{Q_{U}^{N}\left(\mathcal{T}_{\epsilon}^{(N)}\left(Q_{U}\right)\right)}.
\]
%\color{magenta}
(Here $\mathcal{T}_{\epsilon}^{(N)}\left(Q_{U}\right)$ denotes the set of $\epsilon$-typical sequences of length $N$
with respect to the marginal distribution $Q_{U}$.)
%\color{black}
Let $\hat{\mathcal{C}}_{k},k=1,2,...$ be independent copies of $\hat{\mathcal{C}}$.
The codebook $\hat{\mathcal{C}}_{k}$ will be used to generate a nearly
i.i.d. r.v. from the output of $\mathcal{C}_{k}$.
%i.e., the nearly uniform r.v. 
\iffalse
\color{red}
VA comment: Deleted a phrase at the end of the preceding line that did not make sense to me.
\color{black}
\blue
(Lei: Thanks for your revision. Your revision is better.)
\black
\fi
%\color{magenta}
The codebook sequences $\left\{ \mathcal{C}_{k}\right\} ,\left\{ \hat{\mathcal{C}}_{k}\right\} $
are shared by both the terminals.
%\color{black}
We choose the rate $R$ in these two sequence of codebooks
such that 
\[
I_{Q}\left(U;XY\right)<R<H(W)+H_{Q}\left(B|XYU\right).
\]
%\color{magenta}
It can be checked that this is feasible because 
$\left(Q_{U},Q_{B|XU},Q_{Y|BU}\right)$
satisfies the constraints in the expression on the RHS of
\eqref{eq:-18-1-1}. 
%\color{black}

We now describe our scheme in detail. Consider the following sequence
of block codes with each block consisting of $N$ symbols. 
%\blue
For the
first  block (from epoch $1$ to epoch $N$),
the encoder  
%constantly 
sends a sequence of  i.i.d. uniform r.v.'s 
$B_t \sim \mathrm{Unif}(\mathcal{B})$ to the decoder, where $\mathbf{B}_1$ is independent of $\mathbf{X}_1$. 
The decoder generates
$\mathbf{Y}_1$ with a fixed distribution $\hat{Q}_{Y}^{N}$ where $\hat{Q}_{Y}$ is
an optimal distribution attaining $\Delta:=\min_{Q_{Y}}D\left(Q_{Y}\|\pi_{Y|X}|\pi_{X}\right)$. Note that $\Delta$ is finite by assumption. Furthermore, $\mathbf{M}_1,\mathbf{U}_1$ are  set to  be constant.  Obviously, $\mathbf{B}_1, \mathbf{X}_1, \mathbf{Y}_1$ are independent of  $\mathcal{C}_1,\hat{\mathcal{C}}_1$.

For the
$k$-th block 
(from epoch $\left(k-1\right)N+1$ to epoch $kN$) 
with $k\ge 2$, the encoder and decoder adopt the following strategy.
% For the
% $k$-th block {[}from epoch $\left(k-1\right)N+1$ to epoch $kN${]} with $k\ge 2$,
%\black
%For the
%$k$-th block  with $k\ge 2$,  
First the encoder and decoder extract common randomness $\mathbf{M}_{k}$
from the previous block of communication bits $\mathbf{B}_{k-1}$
and common randomness $\mathbf{W}_{k-1}$, by using random binning
based on the codebook $\mathcal{C}_{k}$. That is, the encoder and
decoder generate $\mathbf{M}_{k}=\mathbf{M}\left(\mathbf{B}_{k-1},\mathbf{W}_{k-1}\right)$,
where $\mathbf{M}\left(\mathbf{b},\mathbf{w}\right)$ is the codeword
indexed by $\left(\mathbf{b},\mathbf{w}\right)$ in $\mathcal{C}_{k}$.
Next, the encoder and decoder generate $\mathbf{U}_{k}=\mathbf{U}\left(\mathbf{M}_{k}\right)$
based on the codebook $\hat{\mathcal{C}}_{k}$, where $\mathbf{U}\left(m\right)$
is the codeword indexed by $m$ in $\hat{\mathcal{C}}_{k}$. Then
by using $\left(\mathbf{X}_{k},\mathbf{U}_{k}\right)$, the encoder
generates $\mathbf{B}_{k}$ according to the product conditional distribution
$Q_{B|XU}^{N}$. In fact, the random binning code in the encoder forms
a privacy amplification code with $\left(\mathbf{B}_{k-1},\mathbf{W}_{k-1}\right)$
as public sources and $\left(\mathbf{X}_{k-1},\mathbf{Y}_{k-1},\mathbf{U}_{k-1}\right)$
as private sources. (The target in privacy amplification is to
maximize the alphabet size of the output r.v. $\mathbf{M}_{k}$, generated
from the public sources, under the condition that $\mathbf{M}_{k}$
is nearly uniform and nearly independent of the private sources.)
At the decoder side, upon observing $\left(\mathbf{B}_{k},\mathbf{U}_{k}\right)$
the decoder generates $\mathbf{Y}_{k}$ according to the product conditional
distribution $Q_{Y|BU}^{N}$. Note that this corresponds to the channel
resolvability problem for the channel $Q_{XY|U}$ with $\mathbf{M}_{k}$
considered as the input. (The target in a channel resolvability
problem is 
%to synthesize a target product distribution through a product channel %\color{magenta}
to synthesize a target output distribution of a channel over a block
%\color{black}
%by inputting a uniform r.v., 
%\color{magenta}
by inputting an input block that is a function of a uniform r.v.,
\color{black}
%usually inputting a uniform r.v.
%\color{magenta}
usually one
\color{black}
with the least alphabet size.)  

The distribution for the first $K$
blocks in this code can be expressed as 
%\blue
\begin{align}
P_{\mathcal{C}^{K}\hat{\mathcal{C}}^{K}\mathbf{W}^{K}\mathbf{M}^{K}\mathbf{U}^{K}\mathbf{X}^{K}\mathbf{B}^{K}\mathbf{Y}^{K}}=P_{\mathcal{C}}^{K}P_{\hat{\mathcal{C}}}^{K}P_{W}^{KN}\pi_{X}^{KN} (P_{\mathbf{M}_{1}}P_{\mathbf{U}_{1}}P_{\mathbf{B}_{1}}\hat{Q}_{Y}^{N}) \prod_{k=2}^{K} (P_{\mathbf{M}_{k}|\mathbf{B}_{k-1}\mathbf{W}_{k-1}\mathcal{C}_{k}}P_{\mathbf{U}_{k}|\mathbf{M}_{k}\hat{\mathcal{C}}_{k}}Q_{B|XU}^{N}Q_{Y|BU}^{N}),\label{eq:dist}
\end{align}
%\black
%where $P_{\mathbf{M}_{1}},P_{\mathbf{U}_{1}},P_{\mathbf{B}_{1}}$ are some Dirac measures, 
%\color{magenta}
where $P_{\mathbf{M}_{1}},P_{\mathbf{U}_{1}}$
are some Dirac measures, 
$P_{\mathbf{B}_{1}}$ is as described above,
%\color{black}
$P_{\mathbf{M}_{k}|\mathbf{B}_{k-1}\mathbf{W}_{k-1}\mathcal{C}_{k}}$
corresponds to the deterministic function $\mathbf{M}_{k}=\mathbf{M}\left(\mathbf{B}_{k-1},\mathbf{W}_{k-1}\right)$
with $\mathbf{M}\left(\mathbf{b},\mathbf{w}\right)$ denoting the
codeword indexed by $\left(\mathbf{b},\mathbf{w}\right)$ in $\mathcal{C}_{k}$,
and $P_{\mathbf{U}_{k}|\mathbf{M}_{k}\hat{\mathcal{C}}_{k}}$ corresponds
to the deterministic function $\mathbf{U}_{k}=\mathbf{U}\left(\mathbf{M}_{k}\right)$
with $\mathbf{U}\left(m\right)$ denoting the codeword indexed by
$m$ in $\hat{\mathcal{C}}_{k}$. 

Although the code above is random (since the codebooks are random),
we next show that for this random code, 
\[
\frac{1}{KN}D\left(P_{\mathbf{Y}^{K}|\mathbf{X}^{K}\mathcal{C}^{K}\hat{\mathcal{C}}^{K}}\|\pi_{Y|X}^{KN}|\pi_{X}^{KN}P_{\mathcal{C}}^{K}P_{\hat{\mathcal{C}}}^{K}\right)\to D\left(Q_{Y|X}\|\pi_{Y|X}|\pi_{X}\right)
\]
%as $K\to\infty$, 
%\color{magenta}
as $K\to\infty$ and $N \to \infty$ along an appropriately chosen sequence,
%\color{black}
which implies that there is a sequence of deterministic
codebooks $({c}^{K},\hat{{c}}^{K})$ satisfying 
\[
\frac{1}{KN}D\left(P_{\mathbf{Y}^{K}|\mathbf{X}^{K},\mathcal{C}^{K}={c}^{K},\hat{\mathcal{C}}^{K}=\hat{{c}}^{K}}\|\pi_{Y|X}^{KN}|\pi_{X}^{KN}\right)\to 
D\left(Q_{Y|X}\|\pi_{Y|X}|\pi_{X}\right),
\]
%\color{magenta}
as $K \to \infty$ and $N \to \infty$
along the same sequence.
%\color{black}

For the random code above, we have the following lemma. 
\begin{lem}
\label{lem:For-this-code,} For the random code above, 
\begin{align}
D\left(P_{\mathbf{M}_{k}|\mathbf{X}_{k-1}\mathbf{Y}_{k-1}\mathbf{U}_{k-1}\mathcal{C}_{k}}\|\mathrm{Unif}\left[1:e^{NR}\right]|P_{\mathbf{X}_{k-1}\mathbf{Y}_{k-1}\mathbf{U}_{k-1}}P_{\mathcal{C}_{k}}\right) & \to0\label{eq:D_M}\\
D\left(P_{\mathbf{Y}_{k}|\mathbf{X}_{k}\mathcal{C}^{k}\hat{\mathcal{C}}^{k}}\|Q_{Y|X}^{N}|\pi_{X}^{N}P_{\mathcal{C}}^{k}P_{\hat{\mathcal{C}}}^{k}\right) & \to0\label{eq:D_Y}
\end{align}
uniformly for all 
%\blue 
$k\ge 2$ 
%\black 
as $N\to\infty$. 
\end{lem}
The proof of this lemma is given in Appendix \ref{sec:Proof-of-Lemma}.

% Given the codebooks $\mathcal{C}^{K} = {c}^{K}, \hat{\mathcal{C}}^{K} =\hat{ {c}}^{K}$, 
For the first $K$ blocks induced by the code above we have
\begin{align}
 & D\left(P_{\mathbf{Y}^{K}|\mathbf{X}^{K}\mathcal{C}^{K}\hat{\mathcal{C}}^{K}}\|\pi_{Y|X}^{KN}|\pi_{X}^{KN}P_{\mathcal{C}}^{K}P_{\hat{\mathcal{C}}}^{K}\right)\nonumber \\
 & =\sum_{\mathbf{x}^{K},\mathbf{y}^{K},c^{K},\hat{c}^{K}}P(c^{K})P(\hat{c}^{K})\pi(\mathbf{x}^{K})P(\mathbf{y}^{K}|\mathbf{x}^{K},c^{K},\hat{c}^{K})\log\frac{\prod_{k=1}^{K}P(\mathbf{y}_{k}|\mathbf{x}^{k},\mathbf{y}^{k-1},c^{k},\hat{c}^{k})}{\pi(\mathbf{y}^{K}|\mathbf{x}^{K})}\label{eq:-27}\\
 & =\sum_{k=1}^{K}\sum_{\mathbf{x}^{k},\mathbf{y}^{k},c^{k},\hat{c}^{k}}P(c^{k})P(\hat{c}^{k})\pi(\mathbf{x}^{k}) 
 %\blue 
 P(\mathbf{y}^{k}|\mathbf{x}^{k},c^{k},\hat{c}^{k})
 %\black
 \left(\log\frac{P(\mathbf{y}_{k}|\mathbf{x}^{k},\mathbf{y}^{k-1},c^{k},\hat{c}^{k})}{P(\mathbf{y}_{k}|\mathbf{x}_{k},c^{k},\hat{c}^{k})}+\log\frac{P(\mathbf{y}_{k}|\mathbf{x}_{k},c^{k},\hat{c}^{k})}{\pi(\mathbf{y}_{k}|\mathbf{x}_{k})}\right)\\
 & =\sum_{k=1}^{K}I\left(\mathbf{Y}_{k};\mathbf{X}^{k-1}\mathbf{Y}^{k-1}|\mathbf{X}_{k}\mathcal{C}^{k}\hat{\mathcal{C}}^{k}\right)+\sum_{k=1}^{K}D\left(P_{\mathbf{Y}_{k}|\mathbf{X}_{k}\mathcal{C}^{k}\hat{\mathcal{C}}^{k}}\|\pi_{Y|X}^{N}|\pi_{X}^{N}P_{\mathcal{C}}^{k}P_{\hat{\mathcal{C}}}^{k}\right),\label{eq:-19}
\end{align}
where \eqref{eq:-27} follows since in our scheme, $\mathbf{Y}_{k}\leftrightarrow(\mathbf{X}^{k},\mathbf{Y}^{k-1},\mathcal{C}^{k},\hat{\mathcal{C}}^{k})\leftrightarrow(\mathbf{X}_{k+1}^{K},\mathcal{C}_{k+1}^{K},\hat{\mathcal{C}}_{k+1}^{K})$
under the distribution $P$ (which can be easily seen from the expression
of the joint distribution in \eqref{eq:dist}).

We first consider the first term in \eqref{eq:-19} for $k \ge 2$. From the expression
of the joint distribution in \eqref{eq:dist}, we have that for the
considered code, $\left(\mathbf{X}^{k-1},\mathbf{Y}^{k-1}\right)\leftrightarrow(\mathbf{M}_{k},\mathcal{C}^{k},\hat{\mathcal{C}}^{k})\leftrightarrow\left(\mathbf{X}_{k},\mathbf{Y}_{k}\right)$
forms a Markov chain, and so does $\left(\mathbf{X}^{k-2},\mathbf{Y}^{k-2},\mathcal{C}^{k-1},\hat{\mathcal{C}}^{k}\right)\leftrightarrow(\mathbf{U}_{k-1},\mathcal{C}_{k})\leftrightarrow\left(\mathbf{X}_{k-1},\mathbf{Y}_{k-1},\mathbf{M}_{k}\right)$.
More specifically, the second Markov chain follows since 
\begin{align}
P_{\mathbf{W}_{k-1}\mathbf{B}_{k-1}\mathbf{X}_{k-1}\mathbf{Y}_{k-1}\mathbf{M}_{k}|\mathbf{U}_{k-1}\mathbf{X}^{k-2}\mathbf{Y}^{k-2}\mathcal{C}^{k}\hat{\mathcal{C}}^{k}}=P_{W}^{N}\pi_{X}^{N}Q_{B|XU}^{N}Q_{Y|BU}^{N}P_{\mathbf{M}_{k}|\mathbf{B}_{k-1}\mathbf{W}_{k-1}\mathcal{C}_{k}}.\label{eq:dist2}
\end{align}
Hence, 
\begin{align}
I\left(\mathbf{Y}_{k};\mathbf{X}^{k-1}\mathbf{Y}^{k-1}|\mathbf{X}_{k}\mathcal{C}^{k}\hat{\mathcal{C}}^{k}\right) & \leq I\left(\mathbf{X}_{k}\mathbf{Y}_{k};\mathbf{X}^{k-1}\mathbf{Y}^{k-1}|\mathcal{C}^{k}\hat{\mathcal{C}}^{k}\right)\nonumber \\
 & \leq I\left(\mathbf{M}_{k};\mathbf{X}^{k-1}\mathbf{Y}^{k-1}|\mathcal{C}^{k}\hat{\mathcal{C}}^{k}\right)\nonumber \\
 & \leq I\left(\mathbf{M}_{k};\mathbf{X}_{k-1}\mathbf{Y}_{k-1}\mathbf{U}_{k-1}|\mathcal{C}_{k}\right).\label{eq:}
\end{align}

We have that 
\begin{align}
 & D\left(P_{\mathbf{M}_{k}|\mathbf{X}_{k-1}\mathbf{Y}_{k-1}\mathbf{U}_{k-1}\mathcal{C}_{k}}\|\mathrm{Unif}\left[1:e^{NR}\right]|P_{\mathbf{X}_{k-1}\mathbf{Y}_{k-1}\mathbf{U}_{k-1}}P_{\mathcal{C}_{k}}\right)\nonumber \\
 & =I\left(\mathbf{M}_{k};\mathbf{X}_{k-1}\mathbf{Y}_{k-1}\mathbf{U}_{k-1}|\mathcal{C}_{k}\right)+D\left(P_{\mathbf{M}_{k}|\mathcal{C}_{k}}\|\mathrm{Unif}\left[1:e^{NR}\right]|P_{\mathcal{C}_{k}}\right)\\
 & \geq I\left(\mathbf{M}_{k};\mathbf{X}_{k-1}\mathbf{Y}_{k-1}\mathbf{U}_{k-1}|\mathcal{C}_{k}\right),\label{eq:-1}
\end{align}
where the equality follows since $\log \frac{P_{\mathbf{M}_{k}|\mathbf{X}_{k-1}\mathbf{Y}_{k-1}\mathbf{U}_{k-1}\mathcal{C}_{k}}}{\mathrm{Unif}\left[1:e^{NR}\right]}=\log \frac{P_{\mathbf{M}_{k}|\mathbf{X}_{k-1}\mathbf{Y}_{k-1}\mathbf{U}_{k-1}\mathcal{C}_{k}}}{P_{\mathbf{M}_{k}| \mathcal{C}_{k}}}+\log \frac{P_{\mathbf{M}_{k}| \mathcal{C}_{k}}}{\mathrm{Unif}\left[1:e^{NR}\right]}$.
\iffalse
\color{red}
VA comment: It would be useful to give some details for the preceding claim.
\color{black}
\blue 
Lei: Thanks for your suggestion. I have added the explanation for the equality above.
\black
\fi
Hence combining \eqref{eq:D_M}, \eqref{eq:}, and \eqref{eq:-1}, for $k \ge 2$, 
we have $I\left(\mathbf{Y}_{k};\mathbf{X}^{k-1}\mathbf{Y}^{k-1}|\mathbf{X}_{k}\mathcal{C}^{k}\hat{\mathcal{C}}^{k}\right)\to0$.

We next consider the second term in \eqref{eq:-19} for $k \ge 2$. 
\begin{align*}
D\left(P_{\mathbf{Y}_{k}|\mathbf{X}_{k}\mathcal{C}^{k}\hat{\mathcal{C}}^{k}}\|\pi_{Y|X}^{N}|\pi_{X}^{N}P_{\mathcal{C}}^{k}P_{\hat{\mathcal{C}}}^{k}\right) & =D\left(P_{\mathbf{Y}_{k}|\mathbf{X}_{k}\mathcal{C}^{k}\hat{\mathcal{C}}^{k}}\|Q_{Y|X}^{N}|\pi_{X}^{N}P_{\mathcal{C}}^{k}P_{\hat{\mathcal{C}}}^{k}\right)\\
 & \qquad+\sum_{\mathbf{x}_{k},\mathbf{y}_{k},c_{k},\hat{c}_{k}}P(c_{k},\hat{c}_{k})\pi(\mathbf{x}_{k})P(\mathbf{y}_{k}|\mathbf{x}_{k},c^{k},\hat{c}^{k})\log\frac{Q(\mathbf{y}_{k}|\mathbf{x}_{k})}{\pi(\mathbf{y}_{k}|\mathbf{x}_{k})}.
\end{align*}
By \eqref{eq:D_Y}, $D\left(P_{\mathbf{Y}_{k}|\mathbf{X}_{k}\mathcal{C}^{k}\hat{\mathcal{C}}^{k}}\|Q_{Y|X}^{N}|\pi_{X}^{N}P_{\mathcal{C}}^{k}P_{\hat{\mathcal{C}}}^{k}\right)\to0$.
% if $R>I_{Q}\left(U;XY\right)$. 
Denote $J$ as a random time index, which is independent of all other
r.v.'s involved in the system. Observe that $\pi_{X}P_{Y_{J}|X_{J}\mathcal{C}^{k}\hat{\mathcal{C}}^{k}}$
and $\pi_{X}Q_{Y|X}$ are respectively the output distributions of
the channel $\left(\mathbf{X}_{k},\mathbf{Y}_{k}\right)\mapsto\left(X_{J},Y_{J}\right)$
with input distributions $\pi_{X}^{N}P_{\mathbf{Y}_{k}|\mathbf{X}_{k}\mathcal{C}^{k}\hat{\mathcal{C}}^{k}}$
and $\pi_{X}^{N}Q_{Y|X}^{N}$. Hence by the data processing inequality
concerning relative entropy, we have for $k \ge 2$,
\[
D\left(P_{Y_{J}|X_{J}\mathcal{C}^{k}\hat{\mathcal{C}}^{k}}\|Q_{Y|X}|\pi_{X}P_{\mathcal{C}}^{k}P_{\hat{\mathcal{C}}}^{k}\right)\leq D\left(P_{\mathbf{Y}_{k}|\mathbf{X}_{k}\mathcal{C}^{k}\hat{\mathcal{C}}^{k}}\|Q_{Y|X}^{N}|\pi_{X}^{N}P_{\mathcal{C}}^{k}P_{\hat{\mathcal{C}}}^{k}\right)\to0.
\]
By Pinsker's inequality, this further implies that $P_{\mathcal{C}}^{k}P_{\hat{\mathcal{C}}}^{k}\pi_{X}P_{Y_{J}|X_{J}\mathcal{C}^{k}\hat{\mathcal{C}}^{k}}$
converges to $P_{\mathcal{C}}^{k}P_{\hat{\mathcal{C}}}^{k}\pi_{X}Q_{Y|X}$
under the total variation distance, which further implies that 
\begin{align*}
 & \left|\frac{1}{N}\sum_{\mathbf{x}_{k},\mathbf{y}_{k},c_{k},\hat{c}_{k}}P(c_{k},\hat{c}_{k})\pi(\mathbf{x}_{k})P(\mathbf{y}_{k}|\mathbf{x}_{k},c_{k},\hat{c}_{k})\log\frac{Q(\mathbf{y}_{k}|\mathbf{x}_{k})}{\pi(\mathbf{y}_{k}|\mathbf{x}_{k})}-D\left(Q_{Y|X}\|\pi_{Y|X}|\pi_{X}\right)\right|\\
 & =\left|\sum_{x,y,c_{k},\hat{c}_{k}}P(c_{k},\hat{c}_{k})\pi_{X}(x)\left(P_{Y_{J}|X_{J}\mathcal{C}_{k}\hat{\mathcal{C}}_{k}}(y|x,c_{k},\hat{c}_{k})-Q_{Y|X}(y|x)\right)\log\frac{Q_{Y|X}(y|x)}{\pi_{Y|X}(y|x)}\right|\\
 & \leq\sum_{x,y,c_{k},\hat{c}_{k}}P(c_{k},\hat{c}_{k})\pi_{X}(x)\left|P_{Y_{J}|X_{J}\mathcal{C}_{k}\hat{\mathcal{C}}_{k}}(y|x,c_{k},\hat{c}_{k})-Q_{Y|X}(y|x)\right|\left|\log\frac{Q_{Y|X}(y|x)}{\pi_{Y|X}(y|x)}\right|\\
 & \leq\left|P_{\mathcal{C}}^{k}P_{\hat{\mathcal{C}}}^{k}\pi_{X}P_{Y_{J}|X_{J}\mathcal{C}^{k}\hat{\mathcal{C}}^{k}}-P_{\mathcal{C}}^{k}P_{\hat{\mathcal{C}}}^{k}\pi_{X}Q_{Y|X}\right|_{\mathrm{TV}}\times\max_{x,y}\left|\log\frac{Q_{Y|X}(y|x)}{\pi_{Y|X}(y|x)}\right|\\
 & \to0.
\end{align*}
In the last inequality above, the max term is finite since $D\left(Q_{Y|X}\|\pi_{Y|X}|\pi_{X}\right)$
is finite and $\pi_{X}$ is fully supported. Hence, for $k \ge 2$, $\frac{1}{N}D\left(P_{\mathbf{Y}_{k}|\mathbf{X}_{k}\mathcal{C}^{k}\hat{\mathcal{C}}^{k}}\|\pi_{Y|X}^{N}|\pi_{X}^{N}P_{\mathcal{C}}^{k}P_{\hat{\mathcal{C}}}^{k}\right)\to D\left(Q_{Y|X}\|\pi_{Y|X}|\pi_{X}\right)$.

Hence combining the two points above yields for any given $N$ and noting that both the summands for $k=1$ in the first and second summations in \eqref{eq:-19} are finite, we have that  $\frac{1}{KN}D\left(P_{\mathbf{Y}^{K}|\mathbf{X}^{K}\mathcal{C}^{K}\hat{\mathcal{C}}^{K}}\|\pi_{Y|X}^{KN}|\pi_{X}^{KN}P_{\mathcal{C}}^{K}P_{\hat{\mathcal{C}}}^{K}\right)\to D\left(Q_{Y|X}\|\pi_{Y|X}|\pi_{X}\right)$
as $K\to\infty$, which implies that there is a sequence of deterministic
codebooks $({c}^{K},\hat{{c}}^{K})$ satisfying 
\begin{equation}
\frac{1}{KN}D\left(P_{\mathbf{Y}^{K}|\mathbf{X}^{K},\mathcal{C}^{K}={c}^{K},\hat{\mathcal{C}}^{K}=\hat{{c}}^{K}}\|\pi_{Y|X}^{KN}|\pi_{X}^{KN}\right)\to D\left(Q_{Y|X}\|\pi_{Y|X}|\pi_{X}\right).\label{eq:-14}
\end{equation}
Hence,  $\Gamma\left(\pi_{XY},P_{W}\right)\leq\overline{\Delta}\left(\pi_{XY},P_{W}\right)$.

%\blue
We next proceed to prove $\Gamma\left(\pi_{XY},P_{W}\right)\leq\Delta\left(\pi_{XY},P_{W}\right)$. 
% This is because, if we have shown $\Gamma\left(\pi_{XY},P_{W}\right)\leq\overline{\Delta}\left(\pi_{XY},P_{W}\right)$,
% then denoting 
Let $(P_{UV},P_{B|XUV},P_{Y|BUV})$ be a joint distribution such that $H\left(U|V\right) < H\left(BU|XYV\right)+ H(W)$.  
% by time-sharing arguments, we have $\Gamma\left(\pi_{XY},P_{W}\right)\leq\Delta\left(\pi_{XY},P_{W}\right)$.
Denote  $ (Q_{U^m},Q_{B^m|X^m U^m},Q_{Y^m|B^m U^m}):=(P_{U|V}^m(\cdot|v^m),P_{B|XUV}^m(\cdot|\cdot, v^m),P_{Y|BUV}^m(\cdot|\cdot, v^m)) $  for some   $v^m\in \mathcal{T}^{(m)}_{\epsilon'}(P_V),\; \epsilon'>0$. 
We then extend the code above to the $m$-letter version by substituting 
\begin{align}
    (P_W, \pi_X,Q_{U},Q_{B|XU},Q_{Y|BU}) & \leftarrow (P_W^m, \pi_X ^m, Q_{U^m},Q_{B^m|X^m U^m},Q_{Y^m|B^m U^m}) 
\end{align}   
into the code above.
% Note that this  $m$-letter code is applied to a block of $m$-letter source symbols, i.e., $X \leftarrow X^m$. 
In this  $m$-letter code, the basic unit is the supersymbol which consists of  $m$ successive original letters. Even so,  by definition, the random mappings $Q_{B^m|X^m U^m},Q_{Y^m|B^m U^m}$   indeed still work in symbol-by-symbol way, which means that the  $m$-letter code 
is also a feasible code for the single-letter scenario. In other words, the encoder and decoder of   this  $m$-letter code are still a special case of  $(P_{B_{k}|W^{k}X^{k}B^{k-1}},P_{Y_{k}|W^{k}B^{k}Y^{k-1}})$. 
Hence, 
\begin{equation}
\Gamma\left(\pi_{XY},P_{W}\right)\leq  \frac{1}{m} D\left(Q_{Y^m|X^m}\|\pi_{Y|X}^m|\pi_{X}^m\right) 
\end{equation}
as long as $H_Q\left(U^m\right) < H_Q\left(B^m U^m|X^m Y^m \right)+H_Q\left(W^m\right)$. We now claim that this condition for sufficiently large $m$ is in fact equivalent to $H\left(U|V\right) < H\left(BU|XYV\right)+ H(W)$, and moreover, $\frac{1}{m} D\left(Q_{Y^m|X^m}\|\pi_{Y|X}^m|\pi_{X}^m\right)=D\left(P_{Y|XV}\|\pi_{Y|X}|\pi_{X}P_V\right)+o(1)$. We next prove this claim. 
% , where $P_{Y^m|X^m}$ is induced by  
% $P_W^m\pi_X ^m P_{U|V}^m(\cdot|v^m)P_{B|XUV}^m(\cdot|\cdot, v^m)P_{Y|BUV}^m(\cdot|\cdot, v^m)$.

% \inf_{\substack{P_{U},P_{B|XU},P_{Y|BU}:\\
% H\left(U\right) < H\left(BU|XY\right)+H\left(W^m\right)
% }
% }
% \eqref{eq:-18-1-1} for some $v^m\in \mathcal{T}^{(m)}_{\epsilon'}(P_V),\; \epsilon'>0$, 
By the conditional typicality lemma \cite{Gamal},  we have that with high probability, 
\[
(W^m, X^m,U^m,B^m,Y^m) \sim P_W^m \pi_{X}^m P_{U|V}^m(\cdot|v^m)P_{B|XUV}^m(\cdot|\cdot, v^m)P_{Y|BUV}^m(\cdot|\cdot, v^m)
\]
is jointly $\epsilon$-typical with $v^m$ (with respect to the distribution $P_W \pi_X P_V P_{U|V} P_{B|XUV} P_{Y|BUV}$) for some $\epsilon>\epsilon'$ and sufficiently large $m$. Hence, $\frac{1}{m}H(U^m|V^m=v^m) = H(U|V) + o(1),\frac{1}{m}H(B^m U^m|X^m,Y^m,V^m=v^m) = H(
BU|XYV) + o(1), \frac{1}{m}H(W^m|V^m=v^m)=H(W)$, and $\frac{1}{m}D\left(P_{Y|XV}^m(\cdot|\cdot,v^m)\|\pi_{Y|X}^m|\pi_{X}^m\right)=D\left(P_{Y|XV}\|\pi_{Y|X}|\pi_{X}P_V\right)+o(1)$, where $o(1)$ denotes a generic term vanishing as $m\to \infty$.
This implies the claim above.

% Hence, if $\Gamma\left(\pi_{XY},P_{W}\right)\leq\overline{\Delta}\left(\pi_{XY},P_{W}\right)$, then 
By the claim above, we have $\Gamma\left(\pi_{XY},P_{W}\right) \le \lim_{t\uparrow H\left(W\right)} \Psi(t)$. 
Since $H(W) \neq t_{\min}$,  $\Psi(t)$ is continuous at $t=H(W)$. We have   $\Gamma\left(\pi_{XY},P_{W}\right)\leq \Psi(H\left(W\right))=\Delta\left(\pi_{XY},P_{W}\right)$.
% By the convexity, $\Psi(R)$ is continuous on $(R_0,\infty)$ as long as $\Psi(R_0)<+\infty$. 
% By choosing $U,V$ as constants,   $B$ as a  nonconstant r.v., $P_{Y|BUV}=P{Y$
%  $\Gamma\left(\pi_{XY},P_{W}\right)\leq\Delta\left(\pi_{XY},P_{W}\right)$.
This completes the proof of the achievability part.
%\black

\subsection{Converse}

We next consider the converse part. Observe that 
\begin{align*}
D\left(P_{Y^{n}|X^{n}}\|\pi_{Y|X}^{n}|\pi_{X}^{n}\right) & =D\left(\prod_{k=1}^{n}P_{Y_{k}|X^{k}Y^{k-1}}\|\pi_{Y|X}^{n}|\pi_{X}^{n}\right)\\
 & =\sum_{k=1}^{n}D\left(P_{Y_{k}|X^{k}Y^{k-1}}\|\pi_{Y|X}|\pi_{X}^{k}P_{Y^{k-1}|X^{k}}\right).
\end{align*}
Denote $K\sim\mathrm{Unif}\left[1:n\right]$ as a random time index,
which is independent of all other r.v.'s involved in the system. Define
$U:=\left(B^{K-1},W^{K}\right),V:=\left(X^{K-1},Y^{K-1},K\right),B:=B_{K},X:=X_{K},Y:=Y_{K}$.
Then 
\begin{align}
\frac{1}{n}D\left(P_{Y^{n}|X^{n}}\|\pi_{Y|X}^{n}|\pi_{X}^{n}\right) & =D\left(P_{Y|XV}\|\pi_{Y|X}|\pi_{X}P_{V}\right).\label{eq:-8}
\end{align}
It is easy to verify that 
\begin{align*}
P_{UVBXY}(u,v,b,x,y) & =P_{W}^{k}(w^{k})P_{K}(k)P_{X^{k-1}Y^{k-1}|W^{k}}\left(x^{k-1},y^{k-1}|w^{k}\right)P_{B^{k-1}|X^{k-1}Y^{k-1}W^{k}}\left(b^{k-1}|x^{k-1},y^{k-1},w^{k}\right)\\
 & \qquad\times\pi_{X}(x)P_{B_{k}|X^{k}B^{k-1}W^{k}}\left(b_{k}|x^{k},b^{k-1},w^{k}\right)P_{Y_{k}|B^{k}Y^{k-1}W^{k}}\left(y_{k}|b^{k},y^{k-1},w^{k}\right)\\
 & =P_{UV}\left(u,v\right)\pi_{X}(x)P_{B|XUV}\left(b|x,u,v\right)P_{Y|BUV}\left(y|b,u,v\right).
\end{align*}
Hence it remains to show $H\left(U|V\right)\leq H\left(BU|XYV\right)+H\left(W\right)$.
This can be easily verified as follows: 
\begin{align}
 & H\left(BU|XYV\right)-H\left(U|V\right)\nonumber \\
 & =\frac{1}{n}\sum_{k=1}^{n}\left\{ H\left(B^{k}W^{k}|X^{k}Y^{k}\right)-H\left(B^{k-1}W^{k}|X^{k-1}Y^{k-1}\right)\right\} \label{eq:-38}\\
 & =\frac{1}{n}\sum_{k=1}^{n}\left\{ H\left(B^{k}W^{k}|X^{k}Y^{k}\right)-H\left(B^{k-1}W^{k-1}|X^{k-1}Y^{k-1}\right)-H\left(W_{k}|X^{k-1}Y^{k-1}B^{k-1}W^{k-1}\right)\right\} \nonumber \\
 & =\frac{1}{n}H\left(B^{n}W^{n}|X^{n}Y^{n}\right)-H\left(W\right)\label{eq:-25}\\
 & \geq-H\left(W\right),\label{eq:-39}
\end{align}
where \eqref{eq:-25} follows since $W_{k}$ is independent of $X^{k-1}Y^{k-1}B^{k-1}W^{k-1}$
and has entropy $H\left(W\right)$.

\subsection{\label{sec:Proof-of-Lemma}Proof of Lemma \ref{lem:For-this-code,}}

We now prove Lemma \ref{lem:For-this-code,} by using a R\'enyi entropy
method. Recall that the rate $R$ is chosen such that 
\begin{equation}
I_{Q}\left(U;XY\right)<R<H(W)+H_{Q}\left(B|XYU\right).\label{eq:-31}
\end{equation}
The condition can be relaxed to 
\begin{align}
 & \left(1+\epsilon\right)D_{1+s}\left(Q_{XY|U}\|Q_{XY}|Q_{U}\right)\label{eq:-34}\\
 & <R \nonumber\\
 %\label{eq:-32}\\
 & <\left(1-\epsilon\right)\sum_{u}Q_{U}\left(u\right)H_{1+s}\left(B|XY,U=u\right)+H_{1+s}(W)\label{eq:-33}
\end{align}
for some $\epsilon,s>0$, since both the expressions in \eqref{eq:-34}
and \eqref{eq:-33} are continuous in $\epsilon$ and $s$
%\color{magenta}
and we have $H_{Q}\left(B|XYU\right) = 
\sum_{u}Q_{U}\left(u\right)H\left(B|XY,U=u\right)$.
%\color{black}

We first prove that if 
%\color{magenta}
the upper bound on $R$ given by 
%\color{black}
\eqref{eq:-33} holds, then we have \eqref{eq:D_M}.
To show this, we need the following lemma on one-shot privacy
amplification. 
\iffalse
\color{red}
Made the reference more precise in the following lemma. Please check. It also seems that this Lemma may actually be in Reference [18] of 
\cite{Hayashi17}, specifically in Appendix I of that paper.
\color{black}
\blue
Lei: Thanks for your careful checking. I have updated the reference.
\black 
\fi
\begin{lem}
\cite[Equation (29)]{Hayashi11}
\label{lem:oneshotach-1} Consider a random mapping
$f_{\mathcal{C}}:\calX\rightarrow\calM:=\{1,\ldots,e^{R}\}$. We set
$\mathcal{C}=\left\{ M\left(x\right)\right\} _{x\in\calX}$ with $M\left(x\right),x\in\calX$
drawn independently for different $x$'s and according to the uniform
distribution $\mathrm{Unif}\left[1:e^{R}\right]$, and set $f_{\mathcal{C}}\left(x\right)=M\left(x\right)$.
This forms a random binning code. For this random code, we have for
$s\in(0,1]$ and any distribution $P_{XY}$, 
\begin{align}
 & e^{sD_{1+s}(P_{f_{\mathcal{C}}\left(X\right)|Y\mathcal{C}}\|\mathrm{Unif}\left[1:e^{R}\right]|P_{Y}P_{\mathcal{C}})}\nonumber \\
 & \leq1+e^{-s\left(H_{1+s}\left(X|Y\right)-R\right)}.\label{eq:-123-1}
\end{align}
\end{lem}
Note that the codebook in this lemma is generated in the same way
as the codebook $\mathcal{C}_{k}$ in our scheme. By applying the
lemma above with substitution $X\leftarrow(\mathbf{B}_{k-1},\mathbf{W}_{k-1}),Y\leftarrow(\mathbf{X}_{k-1},\mathbf{Y}_{k-1},\mathbf{U}_{k-1})$,
we have 
\begin{align}
 & D_{1+s}\left(P_{\mathbf{M}_{k}|\mathbf{X}_{k-1}\mathbf{Y}_{k-1}\mathbf{U}_{k-1}\mathcal{C}_{k}}\|\mathrm{Unif}\left[1:e^{nR}\right]|P_{\mathbf{X}_{k-1}\mathbf{Y}_{k-1}\mathbf{U}_{k-1}}P_{\mathcal{C}_{k}}\right)\nonumber \\
 & \leq\frac{1}{s}\log\left[1+e^{-s\left(H_{1+s}\left(\mathbf{B}_{k-1}\mathbf{W}_{k-1}|\mathbf{X}_{k-1}\mathbf{Y}_{k-1}\mathbf{U}_{k-1}\right)-NR\right)}\right]\nonumber \\
 & \leq\frac{1}{s}e^{-s\left(H_{1+s}\left(\mathbf{B}_{k-1}\mathbf{W}_{k-1}|\mathbf{X}_{k-1}\mathbf{Y}_{k-1}\mathbf{U}_{k-1}\right)-NR\right)}.\label{eq:-36}
\end{align}
Note that $\mathbf{W}_{k-1}$ is in fact independent of $(\mathbf{B}_{k-1},\mathbf{X}_{k-1},\mathbf{Y}_{k-1},\mathbf{U}_{k-1})$,
since in the first $k-1$ blocks, only $\mathbf{W}_{1},\mathbf{W}_{2},...,\mathbf{W}_{k-2}$
are used in the encoding process. Hence, 
\begin{align}
H_{1+s}\left(\mathbf{B}_{k-1}\mathbf{W}_{k-1}|\mathbf{X}_{k-1}\mathbf{Y}_{k-1}\mathbf{U}_{k-1}\right)=H_{1+s}\left(\mathbf{B}_{k-1}|\mathbf{X}_{k-1}\mathbf{Y}_{k-1}\mathbf{U}_{k-1}\right)+NH_{1+s}(W).\label{eq:indpW}
\end{align}
On the other hand, 
%\blue 
for $k\ge 2$, 
%\black
\begin{align}
 & \frac{1}{N}H_{1+s}\left(\mathbf{B}_{k-1}|\mathbf{X}_{k-1}\mathbf{Y}_{k-1}\mathbf{U}_{k-1}\right)\nonumber \\
 & =\frac{1}{sN}\log\left[\mathbb{E}_{\mathbf{U}_{k-1}}\sum_{\mathbf{x},\mathbf{y}}Q_{XY|U}^{N}\left(\mathbf{x},\mathbf{y}|\mathbf{U}_{k-1}\right)\sum_{\mathbf{b}}Q_{B|XYU}^{N}\left(\mathbf{b}|\mathbf{x},\mathbf{y},\mathbf{U}_{k-1}\right)^{1+s}\right]\label{eq:-29}\\
 & =\frac{1}{sN}\log\left[\sum_{m}P_{\mathbf{M}_{k-1}}\left(m\right)\prod_{i=1}^{N}\left(\sum_{x,y}Q_{XY|U}(x,y|U_{i}\left(m\right))\sum_{b}Q_{B|XYU}(b|x,y,U_{i}\left(m\right))^{1+s}\right)\right]\nonumber \\
 & =\frac{1}{sN}\log\left[\sum_{m}P_{\mathbf{M}_{k-1}}\left(m\right)e^{sN\sum_{u}T_{\mathbf{U}\left(m\right)}\left(u\right)H_{1+s}\left(B|XY,U=U_{i}\left(m\right)\right)}\right]\label{eq:typeU}\\
 & \geq\frac{1}{sN}\log\left[\sum_{m}P_{\mathbf{M}_{k-1}}\left(m\right)e^{\left(1-\epsilon\right)sN\sum_{u}Q_{U}\left(u\right)H_{1+s}\left(B|XY,U=u\right)}\right]\label{eq:typeave}\\
 & =\left(1-\epsilon\right)\sum_{u}Q_{U}\left(u\right)H_{1+s}\left(B|XY,U=u\right),\label{eq:-35}
\end{align}
where $T_{\mathbf{U}\left(m\right)}$ in \eqref{eq:typeU} denotes
the empirical distribution of the sequence $\mathbf{U}\left(m\right)$,
and \eqref{eq:typeave} follows by combining the typical average lemma
on p. 26 of \cite{Gamal} and the fact that by the construction of
the codebook, all codewords $\mathbf{U}\left(m\right)$ come from
$\mathcal{T}_{\epsilon}^{(n)}\left(Q_{U}\right)$. 
%\blue
In fact, for $k=2$, \eqref{eq:-35}  still holds since in this case $\mathbf{B}_1$ is uniform and independent of $\mathbf{X}_1,\mathbf{Y}_1$ and $\mathbf{U}_1$ is set to a constant.
%\black
Substituting \eqref{eq:indpW}
and \eqref{eq:-35} into \eqref{eq:-36}, we have \eqref{eq:D_M},
i.e., 
\begin{equation}
D_{1+s}\left(P_{\mathbf{M}_{k}|\mathbf{X}_{k-1}\mathbf{Y}_{k-1}\mathbf{U}_{k-1}\mathcal{C}_{k}}\|\mathrm{Unif}\left[1:e^{NR}\right]|P_{\mathbf{X}_{k-1}\mathbf{Y}_{k-1}\mathbf{U}_{k-1}}P_{\mathcal{C}_{k}}\right)\to0\label{eq:-2}
\end{equation}
uniformly for all $k$ as $N\to\infty$.  

We next prove that if 
%\eqref{eq:-32} and \eqref{eq:-33} hold, 
%\color{magenta}
the inequality in \eqref{eq:-34} holds,
%\color{black}
then
we have \eqref{eq:D_Y}. First, by the data processing inequality,
\begin{align*}
 & D_{1+s}\left(P_{\mathbf{M}_{k}|\mathbf{X}_{k-1}\mathbf{Y}_{k-1}\mathbf{U}_{k-1}\mathcal{C}^{k}\hat{\mathcal{C}}^{k-1}}\|\mathrm{Unif}\left[1:e^{NR}\right]|P_{\mathbf{X}_{k-1}\mathbf{Y}_{k-1}\mathbf{U}_{k-1}}P_{\mathcal{C}}^{k}P_{\hat{\mathcal{C}}}^{k-1}\right)\\
 & \geq D_{1+s}\left(P_{\mathbf{M}_{k}|\mathcal{C}^{k}\hat{\mathcal{C}}^{k-1}}\|\mathrm{Unif}\left[1:e^{NR}\right]|P_{\mathcal{C}}^{k}P_{\hat{\mathcal{C}}}^{k-1}\right).
\end{align*}
In fact, the LHS above is identical to the LHS of \eqref{eq:D_M}
(or \eqref{eq:-2}), since $\left(\mathbf{X}^{k-2},\mathbf{Y}^{k-2},\mathcal{C}^{k-1},\hat{\mathcal{C}}^{k}\right)\leftrightarrow(\mathbf{U}_{k-1},\mathcal{C}_{k})\leftrightarrow\left(\mathbf{X}_{k-1},\mathbf{Y}_{k-1},\mathbf{M}_{k}\right)$
holds under the distribution $P$ (see the reasoning around \eqref{eq:dist2}).
Combining this with \eqref{eq:D_M}, we have 
\begin{align*}
D_{1+s}\left(P_{\mathbf{M}_{k}|\mathcal{C}^{k}\hat{\mathcal{C}}^{k-1}}\|\mathrm{Unif}\left[1:e^{NR}\right]|P_{\mathcal{C}}^{k}P_{\hat{\mathcal{C}}}^{k-1}\right) & \to0
\end{align*}
uniformly for all $k$ as $N\to\infty$. That is, 
\begin{align}
\frac{1}{N}H_{1+s}\left(P_{\mathbf{M}_{k}|\mathcal{C}^{k}\hat{\mathcal{C}}^{k-1}}|P_{\mathcal{C}}^{k}P_{\hat{\mathcal{C}}}^{k-1}\right) & \to R\label{eq:-13}\\
 & >\left(1+\epsilon\right)D_{1+s}\left(Q_{XY|U}\|Q_{XY}|Q_{U}\right).\label{eq:-28}
\end{align}

Now we need the following lemma on one-shot channel resolvability. 
%\color{magenta}
This can be proved by a technique similar to that used in Lemma \ref{lem:oneshotach-2}, for which we have given a complete proof.
%\color{black}
\iffalse
\color{red}
Please give a precise location in 
\cite{yu2019renyi} for the following Lemma.

\blue 
Lei: Have updated the reference. 
\color{black}
\fi
\begin{lem}
\cite[Lemma 1]{yu2019renyi}\label{lem:oneshotach} Consider a random mapping
$f_{\mathcal{C}}:\calW\rightarrow\calX$. We set $\mathcal{C}=\left\{ X\left(w\right)\right\} _{w\in\calW}$
with $X\left(w\right),w\in\calW$ drawn independently for different
$w$'s and according to a same distribution $P_{X}$, and set $f_{\mathcal{C}}\left(w\right)=X\left(w\right)$.
This forms a random code. For this random code, we have for $s\in(0,1]$
and any distributions $P_{W},P_{Y|X}$ and $Q_{Y}$, 
\begin{align}
 & e^{sD_{1+s}(P_{Y|\mathcal{C}}\|Q_{Y}|P_{\mathcal{C}})}\nonumber \\
 & \leq e^{sD_{1+s}\left(P_{Y|X}\|Q_{Y}|P_{X}\right)-sH_{1+s}\left(P_{W}\right)}+e^{sD_{1+s}(P_{Y}\|Q_{Y})},\label{eq:-123}
\end{align}
where the distribution $P_{Y|\mathcal{C}}$ is induced by the ``true''
joint distribution $P_{\mathcal{C}}P_{W}P_{Y|X=f_{\mathcal{C}}(W)}$,
and the distribution $P_{Y}$ is induced by the ``ideal'' joint
distribution $P_{X}P_{Y|X}$. 
\end{lem}
This lemma immediately implies the following conditional version. 
\begin{lem}
\label{lem:oneshotach2} Under the same assumptions as in Lemma \ref{lem:oneshotach},
% Consider a random mapping
% $f_{\mathcal{C}}:\calW\rightarrow\calX$. We set $\mathcal{C}=\left\{ X\left(w\right)\right\} _{w\in\calW}$
% with $X\left(w\right),w\in\calW$ drawn independently for different
% $w$'s and according to a same distribution $P_{X}$, and set $f_{\mathcal{C}}\left(w\right)=X\left(w\right)$.
% This forms a random code. For this random code,
 for $s\in(0,1]$ and any distributions $P_{AW}P_{B},P_{Y|XB}$ and
$Q_{Y|B}$, we have 
\begin{align}
 & e^{sD_{1+s}(P_{Y|AB\mathcal{C}}\|Q_{Y|B}|P_{A}P_{B}P_{\mathcal{C}})}\nonumber \\
 & \leq e^{sD_{1+s}\left(P_{Y|XB}\|Q_{Y|B}|P_{X}P_{B}\right)-sH_{1+s}\left(P_{W|A}|P_{A}\right)}+e^{sD_{1+s}(P_{Y|B}\|Q_{Y|B}|P_{B})},\label{eq:-124}
\end{align}
where the distribution $P_{Y|AB\mathcal{C}}$ is induced by the ``true''
joint distribution $P_{\mathcal{C}}P_{AW}P_{B}P_{Y|B,X=f_{\mathcal{C}}(W)}$,
and the distribution $P_{Y|B}$ is induced by the ``ideal'' joint
distribution $P_{B}P_{X}P_{Y|XB}$. 
\end{lem}
% \begin{rem}
% Although in this conditional version, we set $W,Z$ to be independent under the distribution $P$, this is in fact necessary. More precisely, we can assume $(W,Z) \sim P_{WZ}$ for some joint distribution $P_{WZ}$ in the lemma above, but in which case, the codebook 
% \end{rem}

\begin{IEEEproof}[Proof of Lemma \ref{lem:oneshotach2}]
Applying Lemma \ref{lem:oneshotach} with substitution $P_{W}\leftarrow P_{W|A=a},P_{Y|X}\leftarrow P_{Y|X,B=b}$
and $Q_{Y}\leftarrow Q_{Y|B=b}$, we obtain that 
\begin{align}
 & e^{sD_{1+s}(P_{Y|A=a,B=b,\mathcal{C}}\|Q_{Y|B=b}|P_{\mathcal{C}})}\nonumber \\
 & \leq e^{sD_{1+s}\left(P_{Y|X,B=b}\|Q_{Y|B=b}|P_{X}\right)-sH_{1+s}\left(P_{W|A=a}\right)}+e^{sD_{1+s}(P_{Y|B=b}\|Q_{Y|B=b})}.
\end{align}
Taking expectation with respect to $(A,B)\sim P_{A}P_{B}$ for the
two sides above, we obtain \eqref{eq:-124}. 
\end{IEEEproof}
Recall that 
\[
\widetilde{Q}_{\mathbf{U}}=\frac{Q_{U}^{N}1_{\mathcal{T}_{\epsilon}^{(N)}\left(Q_{U}\right)}}{Q_{U}^{N}\left(\mathcal{T}_{\epsilon}^{(N)}\left(Q_{U}\right)\right)}.
\]
Note that the codebook in Lemmas \ref{lem:oneshotach} and \ref{lem:oneshotach2}
is generated in the same way as the codebook $\hat{\mathcal{C}}_{k}$
in our scheme. Applying Lemma \ref{lem:oneshotach2} with substitution
$A\leftarrow(\mathcal{C}^{k},\hat{\mathcal{C}}^{k-1}),B\leftarrow\mathbf{X}_{k},W\leftarrow\mathbf{M}_{k},X\leftarrow\mathbf{U}_{k},Y\leftarrow\mathbf{Y}_{k},\mathcal{C}\leftarrow\hat{\mathcal{C}}_{k}$
and the corresponding distributions $P_{AW}\leftarrow P_{\mathcal{C}}^{k}P_{\hat{\mathcal{C}}}^{k-1}P_{\mathbf{M}_{k}|\mathcal{C}^{k}\hat{\mathcal{C}}^{k-1}},P_{B}\leftarrow\pi_{X}^{N},P_{X}\leftarrow\widetilde{Q}_{\mathbf{U}},P_{Y|XB}\leftarrow Q_{Y|UX}^{N},Q_{Y|B}\leftarrow Q_{Y|X}^{N}$
(which induces $P_{Y|AB\mathcal{C}}=P_{\mathbf{Y}_{k}|\mathbf{X}_{k}\mathcal{C}^{k}\hat{\mathcal{C}}^{k}}$),
% where $I_{Y|X}$ denotes the identity channel such that the output $Y$ is equal to the input $X$, 
we have 
\begin{align}
 & e^{sD_{1+s}\left(P_{\mathbf{Y}_{k}|\mathbf{X}_{k}\mathcal{C}^{k}\hat{\mathcal{C}}^{k}}\|Q_{Y|X}^{N}|\pi_{X}^{N}P_{\mathcal{C}}^{k}P_{\hat{\mathcal{C}}}^{k}\right)}\nonumber \\
 & \leq e^{sD_{1+s}\left(\widetilde{Q}_{\mathbf{Y}|\mathbf{XU}}\|Q_{Y|X}^{N}|\pi_{X}^{N}\widetilde{Q}_{\mathbf{U}}\right)-sH_{1+s}\left(P_{\mathbf{M}_{k}|\mathcal{C}^{k}\hat{\mathcal{C}}^{k-1}}|P_{\mathcal{C}}^{k}P_{\hat{\mathcal{C}}}^{k-1}\right)}+e^{sD_{1+s}(\widetilde{Q}_{\mathbf{Y}|\mathbf{X}}\|Q_{Y|X}^{N}|\pi_{X}^{N})}.\label{eq:-123-2}
\end{align}
where $\widetilde{Q}_{\mathbf{Y}|\mathbf{XU}}:=Q_{Y|UX}^{N}$ and
$\widetilde{Q}_{\mathbf{Y}|\mathbf{X}}$ are induced by the ``ideal''
joint distribution 
\begin{align}
\widetilde{Q}_{\mathbf{UXY}}:=\widetilde{Q}_{\mathbf{U}}\pi_{X}^{N}Q_{Y|XU}^{N}.\label{eq:Qtilde}
\end{align}
Note that here $P_{\mathcal{C}}^{k}P_{\hat{\mathcal{C}}}^{k-1}P_{\mathbf{M}_{k}|\mathcal{C}^{k}\hat{\mathcal{C}}^{k-1}}$
and $P_{\mathbf{Y}_{k}|\mathbf{X}_{k}\mathcal{C}^{k}\hat{\mathcal{C}}^{k}}$
correspond to the ``true'' conditional distributions induced by
our scheme. Moreover, according to the process of encoding, $\mathbf{X}_{k},\hat{\mathcal{C}}_{k},(\mathcal{C}^{k},\hat{\mathcal{C}}^{k-1},\mathbf{M}_{k})$
are mutually independent. 

On one hand, $\widetilde{Q}_{\mathbf{U}}$ is not far from the product
version $Q_{U}^{N}$, as shown in the following equations: 
\begin{align}
 & D_{1+s}(\widetilde{Q}_{\mathbf{U}}\|Q_{U}^{N})\nonumber \\
 & =\frac{1}{s}\log\sum_{\mathbf{u}}\left(\frac{Q_{U}^{N}\left(\mathbf{u}\right)1\left\{ \mathbf{u}\in\mathcal{T}_{\epsilon}^{(N)}\right\} }{Q_{U}^{N}\left(\mathcal{T}_{\epsilon}^{(N)}\right)}\right)^{1+s}\left(Q_{U}^{N}\left(\mathbf{u}\right)\right)^{-s}\label{eq:-46}\\
 & =\frac{1}{s}\log\sum_{\mathbf{u}\in\mathcal{T}_{\epsilon}^{(N)}}\left(\frac{1}{Q_{U}^{N}\left(\mathcal{T}_{\epsilon}^{(N)}\right)}\right)^{1+s}Q_{U}^{N}\left(\mathbf{u}\right)\\
 & =\log\frac{1}{Q_{U}^{N}\left(\mathcal{T}_{\epsilon}^{(N)}\right)}\label{eq:-52}\\
 & \rightarrow0,\label{eq:-15-2}
\end{align}
where \eqref{eq:-15-2} follows from the fact that $Q_{U}^{n}\left(\mathcal{T}_{\epsilon}^{(N)}\right)\rightarrow1$.
By the data processing inequality 
%\color{magenta}
for R\'{e}nyi divergence
%\color{black}
\cite{Erven} and by the definition
of the distribution $\widetilde{Q}$ in \eqref{eq:Qtilde}, we have
\begin{equation}
D_{1+s}(\widetilde{Q}_{\mathbf{Y}|\mathbf{X}}\|Q_{Y|X}^{N}|\pi_{X}^{N})\leq D_{1+s}(\widetilde{Q}_{\mathbf{YU}|\mathbf{X}}\|Q_{UY|X}^{N}|\pi_{X}^{N})=D_{1+s}(\widetilde{Q}_{\mathbf{U}}\|Q_{U}^{N})
\end{equation}
Hence $D_{1+s}(\widetilde{Q}_{\mathbf{Y}|\mathbf{X}}\|Q_{Y|X}^{N}|\pi_{X}^{N})\rightarrow0$
as well.

On the other hand, by a derivation similar to the steps from \eqref{eq:-29} to
\eqref{eq:-35}, we have 
\begin{align*}
 & \frac{1}{N}D_{1+s}\left(\widetilde{Q}_{\mathbf{Y}|\mathbf{XU}}\|Q_{Y|X}^{N}|\pi_{X}^{N}\widetilde{Q}_{\mathbf{U}}\right)\leq\left(1+\epsilon\right)D_{1+s}\left(Q_{XY|U}\|Q_{XY}|Q_{U}\right),
\end{align*}
%\blue 
since $\widetilde{Q}_{\mathbf{Y}|\mathbf{XU}}=Q_{Y|XU}^{N}$ and any sequences $\mathbf{u}$ such that $\widetilde{Q}_{\mathbf{U}}(\mathbf{u})>0$ have a type close to $Q_U$.
\iffalse
\color{red}
VA comment: It would be useful to have more details for the preceding claim.

\blue
Lei: Have added more information.
\color{black}
\fi

By \eqref{eq:-123-2}, $D_{1+s}\left(P_{\mathbf{Y}_{k}|\mathbf{X}_{k}\mathcal{C}^{k}\hat{\mathcal{C}}^{k}}\|Q_{Y|X}^{N}|\pi_{X}^{N}P_{\mathcal{C}}^{k}P_{\hat{\mathcal{C}}}^{k}\right)\to0$
since the conditions in \eqref{eq:-13} and \eqref{eq:-28} hold.

\section{\label{sec:Proof-of-Theorem-broadcast}Proof of Theorem \ref{thm:sequentialCS-broadcast-1}}
\subsection{Cardinality Bounds}
%\blue 
We first prove the   cardinality bounds  for 
  the calculation of
$\Delta\left(\pi_{XYZ},P_{W}P_{\hat{W}}\right)$.
Note that the constraints in \eqref{eq:-18-1-4-2-1} can be rewritten as 
$H(XYZ|V)-H(BXYZ|UV) \le H(W)$ and $H(XYZ|V)-H(BXYZ|U\hat{U}V) \le H(W)+H(\hat{W})$. By the support lemma in \cite[Appendix C]{Gamal}, the cardinality of $\mathcal{V}$ can be upper bounded by 
$3$, without changing the constraints and the objective function. 
% The joint distribution can be still factorized as   $\pi_{X}P_{UV}P_{B|XUV}P_{Y|BUV}$.
Applying the support lemma in \cite[Appendix C]{Gamal} again, for each $v$, we can restrict the size of the  support  of $P_{U|V=v}$ no larger  than $|\mathcal{X}||\mathcal{Y}||\mathcal{Z}|+1$ 
without changing the linear functionals  $P_{XYZ|V=v}$ and $H(BXYZ|U,V=v),H(BXYZ|\hat{U},U,V=v)$, and hence also without changing the constraints and the objective function. 
Therefore,  the cardinality of $\mathcal{U}$ can be upper bounded by
$3(|\mathcal{X}||\mathcal{Y}||\mathcal{Z}|+1)$.  

Applying the support lemma in \cite[Appendix C]{Gamal} again, for each $(u,v)$, we can restrict the size of the  support  of $P_{\hat{U}|U=u,V=v}$ no larger  than $|\mathcal{B}||\mathcal{X}||\mathcal{Y}||\mathcal{Z}|$ 
without changing the linear functionals  $P_{BXYZ|U=u,V=v}$ and $H(BXYZ|\hat{U},U=u,V=v)$, and hence also without changing the constraints and the objective function (since $P_{XYZ|V=v}$ remains unchanged as well). 
Therefore,  the cardinality of $\hat{\mathcal{U}}$ can be upper bounded by
$3(|\mathcal{X}||\mathcal{Y}||\mathcal{Z}|+1)|\mathcal{B}||\mathcal{X}||\mathcal{Y}||\mathcal{Z}|$.  

We next  prove the   cardinality bounds  for 
  the calculation of
$\hat{\Delta}\left(\pi_{XYZ},P_{W}P_{\hat{W}}\right)$.
The constraints  for this case can be rewritten as 
$H(XYZ|V)-H(XYZ|UV)-H(B|XYZ\hat{U}UV) \le H(W)$ and $H(XYZ|V)-H(BXYZ|\hat{U}UV) \le H(W)+H(\hat{W})$. By the support lemma in \cite[Appendix C]{Gamal},  we can restrict $|\mathcal{V}| \le 3$.  Applying the support lemma in \cite[Appendix C]{Gamal} again, for each $v$, we can restrict the size of the  support  of $P_{U|V=v}$ no larger  than $|\mathcal{X}||\mathcal{Y}||\mathcal{Z}|+1$ 
without changing the linear functionals  $P_{XYZ|V=v}$ and $H(XYZ|U,V=v)-H(B|XYZ\hat{U}U,V=v),H(BXYZ|\hat{U}U,V=v)$, and hence also without changing the constraints and the objective function. 
Therefore,  the cardinality of $\mathcal{U}$ can be upper bounded by
$3(|\mathcal{X}||\mathcal{Y}||\mathcal{Z}|+1)$.  

Applying the support lemma in \cite[Appendix C]{Gamal} again, for each $(u,v)$, we can restrict the size of the  support  of $P_{\hat{U}|U=u,V=v}$ no larger  than $|\mathcal{B}||\mathcal{X}||\mathcal{Y}||\mathcal{Z}|+1$ 
without changing the linear functionals  $P_{BXYZ|U=u,V=v}$ and $H(B|XYZ\hat{U},U=u,V=v),H(BXYZ|\hat{U},U=u,V=v)$, and hence also without changing the constraints and the objective function. 
Therefore,  the cardinality of $\hat{\mathcal{U}}$ can be upper bounded by
$3(|\mathcal{X}||\mathcal{Y}||\mathcal{Z}|+1)(|\mathcal{B}||\mathcal{X}||\mathcal{Y}||\mathcal{Z}|+1)$.  
%\black

\subsection{Upper Bound }

We first prove the upper bound, i.e., $\Gamma\left(\pi_{XYZ},P_{W}P_{\hat{W}}\right)\leq\hat{\Delta}\left(\pi_{XYZ},P_{W}P_{\hat{W}}\right)$,
by using a proof similar to that of Theorem \ref{thm:sequentialCS}.
%\color{magenta}
In order to do this, we prove that
$\Gamma\left(\pi_{XYZ},P_{W}P_{\hat{W}}\right)\leq\Delta^{+}\left(\pi_{XYZ},P_{W}P_{\hat{W}}\right)$,
where $\Delta^{+}\left(\pi_{XYZ},P_{W}P_{\hat{W}}\right)$ is defined like
$\hat{\Delta}\left(\pi_{XYZ},P_{W}P_{\hat{W}}\right)$ except that the 
strict inequalities in the constraints are replaced by weak inequalities and
$\min$ is replaced by $\inf$.
This suffices because under the assumption that 
there is at least one  pair $(y,z)$ such that $\pi_{YZ|X}(y,z|x)>0$ for all $x$ such that $\pi_{X}(x)>0$
we can show that
$\Delta^{+}\left(\pi_{XYZ},P_{W}P_{\hat{W}}\right)$ equals 
$\hat{\Delta}\left(\pi_{XYZ},P_{W}P_{\hat{W}}\right)$ by using an argument similar to that in Lemma \ref{lem:psilem}.
%\color{black}

Let 
%\color{magenta}
$\overline{\Delta}\left(\pi_{XYZ},P_{W}P_{\hat{W}}\right)$
be defined like
$\Delta^{+}\left(\pi_{XYZ},P_{W}P_{\hat{W}}\right)$
but with $V$ replaced with a constant.
%\color{black}
Let
$\left(Q_{U\hat{U}},Q_{B|XU\hat{U}},Q_{Y|BU\hat{U}},Q_{Y|BU}\right)$
be a tuple 
%of optimal distributions that attains the minimum 
%\color{magenta}
that satisfies the constraints 
under the infimum in the definition of
$\overline{\Delta}\left(\pi_{XYZ},P_{W}P_{\hat{W}}\right)$.
%\color{black}
%in the
%definition of $\hat{\Delta}\left(\pi_{XYZ},P_{W}P_{\hat{W}}\right)$
%but with $V$ replaced with a constant.
Let 
\begin{align*}
\mathcal{C} & :=\left\{ \mathbf{M}\left(\mathbf{b},\mathbf{w}\right):\left(\mathbf{b},\mathbf{w}\right)\in\mathcal{B}^{N}\times\mathcal{W}^{N}\right\} \\
\mathcal{C}' & :=\left\{ \hat{\mathbf{M}}\left(\hat{\mathbf{w}}\right):\hat{\mathbf{w}}\in\hat{\mathcal{W}}^{N}\right\} 
\end{align*}
be two random binning codebooks where $\mathbf{M}\left(\mathbf{b},\mathbf{w}\right)\sim\mathrm{Unif}\left[1:e^{NR}\right],\hat{\mathbf{M}}\left(\hat{\mathbf{w}}\right)\sim\mathrm{Unif}\left[1:e^{N\hat{R}}\right]$
are respectively generated independently. Let $\mathcal{C}_{k},k=1,2,...$
be independent copies of $\mathcal{C}$ and $\mathcal{C}_{k}',k=1,2,...$
be independent copies of $\mathcal{C}'$. The codebook sequences $\left\{ \mathcal{C}_{k}\right\} ,\left\{ \mathcal{C}_{k}'\right\} $
are shared by all the terminals, Alice, Bob, and Charles (although
$\left\{ \mathcal{C}_{k}'\right\} $ will not be used by Charles).
Let 
\[
\hat{\mathcal{C}}:=\left\{ \left(\mathbf{U}\left(m\right),\hat{\mathbf{U}}\left(m,\hat{m}\right)\right):m\in\left[1:e^{NR}\right],\hat{m}\in\left[1:e^{N\hat{R}}\right]\right\} 
\]
be another random codebook where $\mathbf{U}\left(m\right)\sim\widetilde{Q}_{\mathbf{U}},\hat{\mathbf{U}}\left(m,\hat{m}\right)\sim\widetilde{Q}_{\hat{\mathbf{U}}|\mathbf{U}}\left(\cdot|\mathbf{U}\left(m\right)\right)$
are generated independently. Here $\widetilde{Q}_{\mathbf{U}}$ and
$\widetilde{Q}_{\hat{\mathbf{U}}|\mathbf{U}}$ are the following truncated
product distributions: 
\begin{align*}
\widetilde{Q}_{\mathbf{U}} & =\frac{Q_{U}^{N}1_{\mathcal{T}_{\epsilon}^{(N)}\left(Q_{U}\right)}}{Q_{U}^{N}\left(\mathcal{T}_{\epsilon}^{(N)}\left(Q_{U}\right)\right)},\\
\widetilde{Q}_{\hat{\mathbf{U}}|\mathbf{U}}\left(\cdot|\mathbf{u}\right) & =\frac{Q_{\hat{U}|U}^{N}\left(\cdot|\mathbf{u}\right)1_{\mathcal{T}_{\epsilon}^{(N)}\left(Q_{U\hat{U}}|\mathbf{u}\right)}}{Q_{\hat{U}|U}^{N}\left(\mathcal{T}_{\epsilon}^{(N)}\left(Q_{U\hat{U}}|\mathbf{u}\right)|\mathbf{u}\right)},\forall\mathbf{u}\in\mathcal{U}^{N}.
\end{align*}
Let $\hat{\mathcal{C}}_{k},k=1,2,...$ be independent copies of $\hat{\mathcal{C}}$.
The codebook sequence $\left\{ \hat{\mathcal{C}}_{k}\right\} $ is
also shared by all the terminals (Alice, Bob, and Charles). We choose
rates $R,\hat{R}$ such that 
\begin{align}
I_{Q}\left(U;XYZ\right) & <R<H(W)+H_{Q}\left(B|XYZU\hat{U}\right),\label{eq:-11}\\
\hat{R} & <H(\hat{W}),\label{eq:-12}\\
I_{Q}\left(U\hat{U};XYZ\right) & <R+\hat{R}.\nonumber 
\end{align}
Such $\left(R,\hat{R}\right)$ exists if and only if 
\begin{align*}
I_{Q}\left(U;XYZ\right) & <H(W)+H_{Q}\left(B|XYZU\hat{U}\right),\\
I_{Q}\left(U\hat{U};XYZ\right) & <H(\hat{W})+H(W)+H_{Q}\left(B|XYZU\hat{U}\right),
\end{align*}
or equivalently, 
\iffalse
\begin{align*}
H\left(U\right) & \le H(W)+H\left(BU|XYZ\right)-I\left(B;\hat{U}|XYZU\right),\\
H\left(U\hat{U}\right) & \le H(\hat{W})+H(W)+H\left(BU\hat{U}|XYZ\right).
\end{align*}
\fi
%\color{magenta}
\begin{align*}
H_{Q}\left(U\right) & < H(W)+H_{Q}\left(BU|XYZ\right)-I_{Q}\left(B;\hat{U}|XYZU\right),\\
H_{Q}\left(U\hat{U}\right) & < H(\hat{W})+H(W)+H_{Q}\left(BU\hat{U}|XYZ\right),
\end{align*}
which are 
satisfied by the tuple
$\left(Q_{U\hat{U}},Q_{B|XU\hat{U}},Q_{Y|BU\hat{U}},Q_{Y|BU}\right)$
by assumption.
%\color{black}

%\blue
Consider the following sequence of superposition codes. 
% We now describe our scheme in detail. Consider the following sequence
% of block codes with each block consisting of $N$ symbols. 
% %\blue
For the
first  block (from epoch $1$ to epoch $N$),
Alice   
%constantly 
sends a sequence of  i.i.d. uniform r.v.'s 
$B_t \sim \mathrm{Unif}(\mathcal{B})$ to Bob and Charles, where $\mathbf{B}_1$ is independent of $\mathbf{X}_1$. 
Bob and Charles respectively generate 
$\mathbf{Y}_1$ with a fixed distribution $\hat{Q}_{Y}^{N}$ and $\mathbf{Z}_1$ with a fixed distribution $\hat{Q}_{Z}^{N}$ where $(\hat{Q}_{Y},\hat{Q}_{Z})$ is
an optimal distribution attaining $\Delta:=\min_{Q_{Y},Q_{Z}}D\left(Q_{Y}Q_{Z}\|\pi_{YZ|X}|\pi_{X}\right)$. Note that $\Delta$ is finite by assumption. Furthermore, $\mathbf{M}_1,\hat{\mathbf{M}}_{1},\mathbf{U}_1,\hat{\mathbf{U}}_{1}$ are  set to  be constant.  Obviously, $\mathbf{B}_1, \mathbf{X}_1, \mathbf{Y}_1$ are independent of  $\mathcal{C}_1,\mathcal{C}_1',\hat{\mathcal{C}}_1$.

For the
$k$-th block 
(from epoch $\left(k-1\right)N+1$ to epoch $kN$) 
with $k\ge 2$, 
the encoder and decoder adopt the following strategy. 
%\black
All the terminals (Alice, Bob, and Charles) extract common
randomness $\mathbf{M}_{k}$ from the previous block of communication
bits $\mathbf{B}_{k-1}$ and common randomness $\mathbf{W}_{k-1}$,
by using random binning based on $\mathcal{C}_{k}$. That is, they
generate $\mathbf{M}_{k}=\mathbf{M}\left(\mathbf{B}_{k-1},\mathbf{W}_{k-1}\right)$,
where $\mathbf{M}\left(\mathbf{b},\mathbf{w}\right)$ is the codeword
indexed by $\left(\mathbf{b},\mathbf{w}\right)$ in $\mathcal{C}_{k}$.
Besides, Alice and Bob also generate $\hat{\mathbf{M}}_{k}=\hat{\mathbf{M}}\left(\hat{\mathbf{W}}_{k}\right)$,
where $\hat{\mathbf{M}}\left(\hat{\mathbf{w}}\right)$ is the codeword
indexed by $\hat{\mathbf{w}}$ in $\mathcal{C}_{k}'$. Next, Alice
and Bob  generate $\left(\mathbf{U}_{k},\hat{\mathbf{U}}_{k}\right)=\left(\mathbf{U}\left(\mathbf{M}_{k}\right),\hat{\mathbf{U}}\left(\mathbf{M}_{k},\hat{\mathbf{M}}_{k}\right)\right)$,
where $\left(\mathbf{U}\left(m\right),\hat{\mathbf{U}}\left(m,\hat{m}\right)\right)$
the codeword indexed by $\left(m,\hat{m}\right)$ in $\hat{\mathcal{C}}_{k}$.
Moreover, $\mathbf{U}_{k}$ is also available at Charles since he
knows $\mathbf{M}_{k}$. Then 
%by using $\left(\mathbf{X}_{k},\mathbf{U}_{k}\right)$,
%color{magenta}
by using $\left(\mathbf{X}_{k},\mathbf{U}_{k},
\hat{\mathbf{U}}_{k}\right)$,
%\color{black}
the encoder Alice generates $\mathbf{B}_{k}$ by the product distribution
$Q_{B|XU\hat{U}}^{N}$. At the decoder sides, upon observing $\left(\mathbf{B}_{k},\mathbf{U}_{k},\hat{\mathbf{U}}_{k}\right)$
Bob generates $\mathbf{Y}_{k}$ by the product distribution $Q_{Y|BU\hat{U}}^{N}$,
and upon observing $\left(\mathbf{B}_{k},\mathbf{U}_{k}\right)$ Charlie
generates $\mathbf{Z}_{k}$ by the product distribution $Q_{Z|BU}^{N}$.
\begin{lem}
\label{lem:For-this-code,-1}For this code, 
\begin{align}
D\left(P_{\mathbf{M}_{k}\hat{\mathbf{M}}_{k}|\mathbf{X}_{k-1}\mathbf{Y}_{k-1}\mathbf{Z}^{k-1}\mathbf{U}_{k-1}\hat{\mathbf{U}}_{k-1}\mathcal{C}_{k}\mathcal{C}_{k}'}\|\mathrm{Unif}\left[1:e^{NR}\right]\mathrm{Unif}\left[1:e^{N\hat{R}}\right]|P_{\mathbf{X}_{k-1}\mathbf{Y}_{k-1}\mathbf{Z}^{k-1}\mathbf{U}_{k-1}\hat{\mathbf{U}}_{k-1}}P_{\mathcal{C}_{k}}P_{\mathcal{C}_{k}'}\right) & \to0\label{eq:D_M-1}\\
D\left(P_{\mathbf{Y}_{k}\mathbf{Z}_{k}|\mathbf{X}_{k}\mathcal{C}^{k}\mathcal{C}^{\prime k}\hat{\mathcal{C}}^{k}}\|Q_{YZ|X}^{N}|\pi_{X}^{N}P_{\mathcal{C}}^{k}P_{\mathcal{C}'}^{k}P_{\hat{\mathcal{C}}}^{k}\right) & \to0\label{eq:D_Y-1}
\end{align}
uniformly for all $k\ge 2$ as $N\to\infty$. 
\end{lem}
%\blue 
The convergence in \eqref{eq:D_M-1} follows since on one hand,  
\[
P_{\mathbf{M}_{k}\hat{\mathbf{M}}_{k}|\mathbf{X}_{k-1}\mathbf{Y}_{k-1}\mathbf{Z}^{k-1}\mathbf{U}_{k-1}\hat{\mathbf{U}}_{k-1}\mathcal{C}_{k}\mathcal{C}_{k}'}=P_{\mathbf{M}_{k}|\mathbf{X}_{k-1}\mathbf{Y}_{k-1}\mathbf{Z}^{k-1}\mathbf{U}_{k-1}\hat{\mathbf{U}}_{k-1}\mathcal{C}_{k}}P_{\hat{\mathbf{M}}_{k}|\mathcal{C}_{k}'}
\]
and hence, the divergence in  \eqref{eq:D_M-1} can be written as 
the sum of the following two divergences 
\begin{align}
&D\left(P_{\mathbf{M}_{k}|\mathbf{X}_{k-1}\mathbf{Y}_{k-1}\mathbf{Z}^{k-1}\mathbf{U}_{k-1}\hat{\mathbf{U}}_{k-1}\mathcal{C}_{k}}\|\mathrm{Unif}\left[1:e^{NR}\right]|P_{\mathbf{X}_{k-1}\mathbf{Y}_{k-1}\mathbf{Z}^{k-1}\mathbf{U}_{k-1}\hat{\mathbf{U}}_{k-1}}P_{\mathcal{C}_{k}}\right) \label{eq:-9}\\
&D\left(P_{\hat{\mathbf{M}}_{k}|\mathcal{C}_{k}'}\|\mathrm{Unif}\left[1:e^{N\hat{R}}\right]|P_{\mathcal{C}_{k}'}\right), \label{eq:-10}
\end{align}
and on the other hand, by Lemma \ref{lem:oneshotach-1}, the divergences in   \eqref{eq:-9} and \eqref{eq:-10} vanish as $N\to \infty$ 
once the upper bounds on $R,\hat{R}$ in \eqref{eq:-11} and \eqref{eq:-12}
hold. 
\iffalse
\color{red}
VA comment: It would be helpful to have more details for the preceding claim.
\blue 
Lei: Have added more information.
\color{black}
\fi

In order to prove \eqref{eq:D_Y-1}, we need the following lemmas,
which are generalizations of Lemma \ref{lem:oneshotach} 
%\color{magenta}
and Lemma \ref{lem:oneshotach2}
respectively
%\color{black}
to superposition
codes. 
\begin{lem}
\label{lem:oneshotach-2} Let $P_{X\hat{X}}$ be a probability distribution. Consider
a random mapping $f_{\mathcal{C}}:\calW\times\hat{\mathcal{W}}\rightarrow\calX\times\hat{\mathcal{X}}$.
We set $\mathcal{C}=\left\{ \left(X\left(w\right),\hat{X}\left(w,\hat{w}\right)\right)\right\} _{w\in\calW}$
with $X\left(w\right),w\in\calW$ drawn independently for different
$w$'s and according to the same distribution $P_{X}$ and given $w$,
$\hat{X}\left(w,\hat{w}\right),\hat{w}\in\hat{\mathcal{W}}$ drawn
independently for different $\hat{w}$'s and according to the same distribution
$P_{\hat{X}|X}\left(\cdot|X\left(w\right)\right)$, and set $f_{\mathcal{C}}\left(w,\hat{w}\right)=\left(X\left(w\right),\hat{X}\left(w,\hat{w}\right)\right)$.
This forms a random superposition code. For this code, we have for
$s\in(0,1]$ and any distributions $P_{W\hat{W}},P_{Y|X\hat{X}}$
and $Q_{Y}$, 
\begin{align}
 & e^{sD_{1+s}(P_{Y|\mathcal{C}}\|Q_{Y}|P_{\mathcal{C}})}\nonumber \\
 & \leq e^{sD_{1+s}\left(P_{Y|X\hat{X}}\|Q_{Y}|P_{X\hat{X}}\right)-sH_{1+s}\left(P_{W\hat{W}}\right)}\nonumber \\
 & \qquad+e^{sD_{1+s}\left(P_{Y|X}\|Q_{Y}|P_{X}\right)-sH_{1+s}\left(P_{W}\right)}+e^{sD_{1+s}(P_{Y}\|Q_{Y})},
\end{align}
where the distribution $P_{Y|\mathcal{C}}$ is induced by the ``true''
joint distribution $P_{\mathcal{C}}P_{W\hat{W}}P_{Y|(X,\hat{X})=f_{\mathcal{C}}(W,\hat{W})}$,
and the distribution $P_{Y}$ is induced by the ``ideal'' joint
distribution $P_{X\hat{X}}P_{Y|X\hat{X}}$. 
\end{lem}
\begin{lem}
\label{lem:oneshotach-2-1} Let $P_{X\hat{X}}$ be a probability distribution.
Consider a random mapping $f_{\mathcal{C}}:\calW\times\hat{\mathcal{W}}\rightarrow\calX\times\hat{\mathcal{X}}$.
We set $\mathcal{C}=\left\{ \left(X\left(w\right),\hat{X}\left(w,\hat{w}\right)\right)\right\} _{w\in\calW}$
with $X\left(w\right),w\in\calW$ drawn independently for different
$w$'s and according to the same distribution $P_{X}$ and given $w$,
$\hat{X}\left(w,\hat{w}\right),\hat{w}\in\hat{\mathcal{W}}$ drawn
independently for different $\hat{w}$'s and according to the same distribution
$P_{\hat{X}|X}\left(\cdot|X\left(w\right)\right)$, and set $f_{\mathcal{C}}\left(w\right)=\left(X\left(w\right),\hat{X}\left(w,\hat{w}\right)\right)$.
This forms a random superposition code. For this code, we have for
$s\in(0,1]$ and any distributions $P_{AW\hat{W}}P_{B},P_{Y|X\hat{X}B}$,
and $Q_{Y|B}$, 
\begin{align}
 & e^{sD_{1+s}(P_{Y|AB\mathcal{C}}\|Q_{Y|B}|P_{A}P_{B}P_{\mathcal{C}})}\nonumber \\
 & \leq e^{sD_{1+s}\left(P_{Y|X\hat{X}B}\|Q_{Y|B}|P_{X\hat{X}}P_{B}\right)-sH_{1+s}\left(P_{W\hat{W}|A}|P_{A}\right)}\nonumber \\
 & \qquad+e^{sD_{1+s}\left(P_{Y|XB}\|Q_{Y|B}|P_{X}P_{B}\right)-sH_{1+s}\left(P_{W|A}|P_{A}\right)}+e^{sD_{1+s}(P_{Y|B}\|Q_{Y|B}|P_{B})},
\end{align}
where the distribution $P_{Y|AB\mathcal{C}}$ is induced by the ``true''
joint distribution $P_{\mathcal{C}}P_{AW\hat{W}}P_{B}P_{Y|B,(X,\hat{X})=f_{\mathcal{C}}(W,\hat{W})}$,
and the distributions $P_{Y|B}$ and $P_{Y|XB}$ are induced by the
``ideal'' joint distribution $P_{B}P_{X\hat{X}}P_{Y|X\hat{X}B}$. 

\end{lem}
Lemma \ref{lem:oneshotach-2-1} can be seen as a conditional version of Lemma \ref{lem:oneshotach-2}. The proof of Lemma \ref{lem:oneshotach-2} is provided in Appendix
\ref{sec:Proof-of-Lemma-1}. The extension of Lemma \ref{lem:oneshotach-2}
to Lemma \ref{lem:oneshotach-2-1} follows similarly to the extension
of Lemma \ref{lem:oneshotach} to Lemma \ref{lem:oneshotach2}. 

By proof steps similar to that of Lemma \ref{lem:For-this-code,}
except for replacing Lemma \ref{lem:oneshotach2} with Lemma \ref{lem:oneshotach-2-1},
one can prove \eqref{eq:D_Y-1}. Specifically, consider the following
substitution in Lemma \ref{lem:oneshotach-2-1}: $A\leftarrow(\mathcal{C}^{k},\mathcal{C}^{\prime k},\hat{\mathcal{C}}^{k-1}),B\leftarrow\mathbf{X}_{k},W\leftarrow\mathbf{M}_{k},\hat{W}\leftarrow\hat{\mathbf{M}}_{k},X\leftarrow\mathbf{U}_{k},\hat{X}\leftarrow\hat{\mathbf{U}}_{k},Y\leftarrow(\mathbf{Y}_{k},\mathbf{Z}_{k}),\mathcal{C}\leftarrow\hat{\mathcal{C}}_{k}$
and the corresponding distributions $P_{AW\hat{W}}\leftarrow P_{\mathcal{C}}^{k}P_{\mathcal{C}'}^{k}P_{\hat{\mathcal{C}}}^{k-1}P_{\mathbf{M}_{k}\hat{\mathbf{M}}_{k}|\mathcal{C}^{k}\hat{\mathcal{C}}^{k-1}},P_{B}\leftarrow\pi_{X}^{N},P_{X}\leftarrow\widetilde{Q}_{\mathbf{U}},P_{\hat{X}|X}\leftarrow\widetilde{Q}_{\hat{\mathbf{U}}|\mathbf{U}},P_{Y|X\hat{X}B}\leftarrow Q_{Y|U\hat{U}X}^{N}Q_{Z|UX}^{N},Q_{Y|B}\leftarrow Q_{YZ|X}^{N}$.
Furthermore, by proof steps similar to those from \eqref{eq:-27}
to \eqref{eq:-14}, one can show 
%\color{magenta}
that $\Gamma\left(\pi_{XYZ},P_{W}P_{\hat{W}}\right) \le \overline{\Delta}\left(\pi_{XYZ},P_{W}P_{\hat{W}}\right)$. The random variable $V$ can be added by an argument similar to that 
given at the end of achievability proof of Theorem \ref{thm:sequentialCS} to conclude that
$\Gamma\left(\pi_{XYZ},P_{W}P_{\hat{W}}\right) \le \Delta^{+}\left(\pi_{XYZ},P_{W}P_{\hat{W}}\right)$.
Since $\Delta^{+}\left(\pi_{XYZ},P_{W}P_{\hat{W}}\right) = \Delta\left(\pi_{XYZ},P_{W}P_{\hat{W}}\right)$ under our assumptions, this 
completes the proof of
%\color{black}
the achievability part of Theorem
\ref{thm:sequentialCS-broadcast-1}.  Here we omit the detailed proofs.

\subsection{Lower Bound }

The lower bound follows similarly to the converse in Theorem \ref{thm:sequentialCS}.
Denote $K\sim\mathrm{Unif}\left[1:n\right]$ as a random time index,
which is independent of all other r.v.'s involved in the system. Define
$U:=\left(B^{K-1},W^{K}\right),\hat{U}:=\hat{W}^{K},V:=\left(X^{K-1},Y^{K-1},Z^{K-1},K\right),B:=B_{K},X:=X_{K},Y:=Y_{K},Z:=Z_{K}$.
Then, following derivations similar to the ones for the converse of
Theorem \ref{thm:sequentialCS}, we have 
\begin{align*}
\frac{1}{n}D\left(P_{Y^{n}Z^{n}|X^{n}}\|\pi_{YZ|X}^{n}|\pi_{X}^{n}\right) & =D\left(P_{YZ|XV}\|\pi_{YZ|X}|\pi_{X}P_{V}\right),\\
H\left(U|V\right) & \le H(W)+H\left(BU|XYZV\right),\\
H\left(U\hat{U}|V\right) & \le H(\hat{W})+H(W)+H\left(BU\hat{U}|XYZV\right).
\end{align*}
Moreover, 
\begin{align*}
P_{U\hat{U}VBXYZ}(u,\hat{u},v,b,x,y,z) & =P_{W}^{k}(w^{k})P_{\hat{W}}^{k}(\hat{w}^{k})P_{K}(k)P\left(b^{k-1},x^{k-1},y^{k-1},z^{k-1}|w^{k},\hat{w}^{k}\right)\\
 & \qquad\times\pi_{X}(x)P\left(b_{k}|x^{k},b^{k-1},w^{k},\hat{w}^{k}\right)P\left(y_{k}|b^{k},y^{k-1},w^{k},\hat{w}^{k}\right)P\left(z_{k}|b^{k},z^{k-1},w^{k}\right)\\
 & =P_{U\hat{U}V}\left(u,v\right)\pi_{X}(x)P_{B|XU\hat{U}V}\left(b|x,u,\hat{u},v\right)P_{Y|BU\hat{U}V}\left(y|b,u,\hat{u},v\right)P_{Z|BUV}\left(z|b,u,v\right).
\end{align*}
Combining all the above yields the lower bound $\Delta\left(\pi_{XYZ},P_{W}P_{\hat{W}}\right)$.

\subsection{\label{sec:Proof-of-Lemma-1}Proof of Lemma \ref{lem:oneshotach-2}}

Observe that 
\begin{align}
 & e^{sD_{1+s}(P_{Y\mathcal{C}}\|Q_{Y}\times P_{\mathcal{C}})}\nonumber \\
 & =\mathbb{E}_{\mathcal{C}}\sum_{y}P^{1+s}\left(y|\mathcal{C}\right)Q^{-s}\left(y\right)\\
 & =\mathbb{E}_{\mathcal{C}}\sum_{y}\sum_{w,\hat{w}}P\left(w,\hat{w}\right)P\left(y|f_{\mathcal{C}}\left(w,\hat{w}\right)\right)\biggl(P\left(w,\hat{w}\right)P\left(y|f_{\mathcal{C}}\left(w,\hat{w}\right)\right)\nonumber \\
 & \qquad+\sum_{\hat{w}'\neq\hat{w}}P(w,\hat{w}')P\left(y|f_{\mathcal{C}}\left(w,\hat{w}'\right)\right)+\sum_{w'\neq w}\sum_{\hat{w}'}P(w',\hat{w}')P\left(y|f_{\mathcal{C}}\left(w',\hat{w}'\right)\right)\biggr)^{s}Q^{-s}\left(y\right)\label{eq:-149}
\end{align}
\iffalse
\color{red}
VA comment: Deleted the reference in the sentence that follows because the inequality is trivial.

\blue 
Lei: Your revision is fine.
\color{black}
\fi
Then using the inequality $(a+b+c)^{s}\le a^{s}+b^{s}+c^{s}$ for
$a,b,c\ge0$ and $0<s\le1$ 
%(see, e.g., \cite[Lemma 6]{yu2019renyi}),
we get 
\begin{align}
 & e^{sD_{1+s}(P_{Y\mathcal{C}}\|Q_{Y}\times P_{\mathcal{C}})}\leq L_{1}+L_{2}+L_{3},\label{eq:-150}
\end{align}
where 
\begin{align}
 & L_{1}:=\sum_{y}\sum_{w,\hat{w}}P^{1+s}\left(w,\hat{w}\right)\mathbb{E}_{\mathcal{C}}\left[P^{1+s}\left(y|f_{\mathcal{C}}\left(w,\hat{w}\right)\right)\right]Q^{-s}\left(y\right)\\
 & L_{2}:=\mathbb{E}_{\mathcal{C}}\sum_{y}\sum_{w,\hat{w}}P\left(w,\hat{w}\right)P\left(y|f_{\mathcal{C}}\left(w,\hat{w}\right)\right)\\
 & \qquad\times\left(\sum_{\hat{w}'\neq\hat{w}}P(w,\hat{w}')P\left(y|f_{\mathcal{C}}\left(w,\hat{w}'\right)\right)\right)^{s}Q^{-s}\left(y\right)\\
 & L_{3}:=\mathbb{E}_{\mathcal{C}}\sum_{y}\sum_{w,\hat{w}}P\left(w,\hat{w}\right)P\left(y|f_{\mathcal{C}}\left(w,\hat{w}\right)\right)\nonumber \\
 & \qquad\times\left(\sum_{w'\neq w}\sum_{\hat{w}'}P(w',\hat{w}')P\left(y|f_{\mathcal{C}}\left(w',\hat{w}'\right)\right)\right)^{s}Q^{-s}\left(y\right).
\end{align}

Furthermore, $L_{1},L_{2}$, and $L_{3}$ can be respectively expressed
or upper bounded as follows. 
\begin{align}
L_{1} & =\sum_{y}\sum_{w,\hat{w}}P^{1+s}\left(w,\hat{w}\right)\sum_{x,\hat{x}}P\left(x,\hat{x}\right)P^{1+s}\left(y|x,\hat{x}\right)Q^{-s}\left(y\right)\\
 & =e^{sD_{1+s}\left(P_{Y|X\hat{X}}\|Q_{Y}|P_{X\hat{X}}\right)-sH_{1+s}\left(W\hat{W}\right)},\label{eq:-3-3-1}
\end{align}
%\color{magenta}
\begin{align}
L_{2} & =\sum_{y}\sum_{w,\hat{w}}P\left(w,\hat{w}\right)\mathbb{E}_{X\left(w\right)}\mathbb{E}_{\hat{X}\left(w,\hat{w}\right)}\left[P_{Y|X\hat{X}}\left(y|X\left(w\right),\hat{X}\left(w,\hat{w}\right)\right)\right]\nonumber \\
 & \qquad\times\mathbb{E}_{\left\{ \hat{X}\left(w,\hat{w}'\right):\hat{w}'\neq\hat{w}\right\} }\left(\sum_{\hat{w}'\neq\hat{w}}P(w,\hat{w}')P_{Y|X\hat{X}}\left(y|X\left(w\right),\hat{X}\left(w,\hat{w}'\right)\right)\right)^{s}Q^{-s}\left(y\right)\label{eq:-114-1-1}\\
 & \leq\sum_{y}\sum_{w,\hat{w}}P\left(w,\hat{w}\right)\mathbb{E}_{X\left(w\right)}\sum_{\hat{x}}P_{\hat{X}|X}\left(\hat{x}|X\left(w\right)\right)P_{Y|X\hat{X}}\left(y|X\left(w\right),\hat{x}\right)\\
 & \qquad\times\left(\sum_{\hat{w}'}P(w,\hat{w}') \mathbb{E}_{  \hat{X}\left(w,\hat{w}'\right)  } P_{Y|X\hat{X}}\left(y|X\left(w\right),\hat{X}\left(w,\hat{w}'\right)\right)\right)^{s}Q^{-s}\left(y\right)\\
 & = \sum_{y}\sum_{w,\hat{w}}P\left(w,\hat{w}\right)\mathbb{E}_{X\left(w\right)}P_{Y|X}\left(y|X\left(w\right)\right)\nonumber \\
 & \qquad\times\left(P(w)P_{Y|X}\left(y|X\left(w\right)\right)\right)^{s}Q^{-s}\left(y\right)\label{eq:-42-2-1}\\
 & =\sum_{w}P\left(w\right)^{1+s}\sum_{y}\sum_{x}P\left(x\right)P\left(y|x\right)^{1+s}Q^{-s}\left(y\right)\nonumber \\
 & =e^{sD_{1+s}\left(P_{Y|X}\|Q_{Y}|P_{X}\right)-sH_{1+s}\left(W\right)},\label{eq:-148-1}
\end{align}
%\color{black}
\iffalse
\color{red}
VA comments: Regarding the calculation for $L_{2}$ above, (1) moved the expectation inside the summation in the fourth line since the random variable over which the expectation is being taken depends on $\hat{w}'$;
(2) in the fifth line I think there is equality.
Please check.
\blue 
Lei: Your revision is fine.
\color{black}
\fi
and 
%\color{magenta}
\begin{align}
L_{3} & = \sum_{y}\sum_{w,\hat{w}}P\left(w,\hat{w}\right)\mathbb{E}_{\mathcal{C}}\left[P\left(y|f_{\mathcal{C}}\left(w,\hat{w}\right)\right)\right]\nonumber \\
 & \qquad\times\mathbb{E}_{\mathcal{C}}\left[\left(\sum_{w'\neq w}\sum_{\hat{w}'}P(w',\hat{w}')P\left(y|f_{\mathcal{C}}\left(w',\hat{w}'\right)\right)\right)^{s}\right]Q^{-s}\left(y\right)\label{eq:-114-1}\\
 & \leq\sum_{y}\sum_{w,\hat{w}}P\left(w,\hat{w}\right)\mathbb{E}_{\mathcal{C}}\left[P\left(y|f_{\mathcal{C}}\left(w,\hat{w}\right)\right)\right]\nonumber \\
 & \qquad\times\left(\sum_{w',\hat{w}'}P(w',\hat{w}')\mathbb{E}_{\mathcal{C}}\left[P\left(y|f_{\mathcal{C}}\left(w',\hat{w}'\right)\right)\right]\right)^{s}Q^{-s}\left(y\right)\label{eq:-42-2}\\
 & =\sum_{y}\sum_{x,\hat{x}}P\left(x,\hat{x}\right)P\left(y|x,\hat{x}\right)\nonumber \\
 & \qquad\times\left(\sum_{x,\hat{x}}P\left(x,\hat{x}\right)P\left(y|x,\hat{x}\right)\right)^{s}Q^{-s}\left(y\right)\\
 & =\sum_{y}P^{1+s}\left(y\right)Q^{-s}\left(y\right)\\
 & =e^{sD_{1+s}(P_{Y}\|Q_{Y})}.\label{eq:-148}
\end{align}
%\color{black}
\iffalse
\color{red}
VA comment: Regarding the calculation for $L_{3}$ I think there is equality in the first line. Please check. 
\blue
Lei: Yes, thanks for careful checking. 
\color{black}
\fi
where \eqref{eq:-42-2} follows since $x\mapsto x^{s}$ is a concave
function 
%\color{magenta}
for $0 < s \le 1$
%\color{black}
and 
%\color{magenta}
we
%\color{black}
relax 
the summation $\sum_{w'\neq w}$ to $\sum_{w'}$.

\section{\label{sec:Proof-of-Theorem-interaction}Proof of Theorem \ref{thm:sequentialCS-interaction}}
%\blue 
The proof of the cardinality bound is similar to the one for Theorem \ref{thm:sequentialCS}. We next prove the equality in \eqref{eq:-18-1-4-1}.
%\black 

\subsection{Achievability}

We first prove the achievability part, i.e., $\Gamma\left(\pi_{SXYZ},P_{W}\right)\leq\Delta\left(\pi_{SXYZ},P_{W}\right)$.
We first prove that 
\iffalse
\begin{equation}
\Gamma\left(\pi_{SXYZ},P_{W}\right)\leq\overline{\Delta}\left(\pi_{SXYZ},P_{W}\right):=\min_{\substack{P_{U},P_{A|SU},P_{B|XU},P_{Y|ABU},P_{Z|ABU}:\\
H\left(U\right)\leq H\left(ABU|SXYZ\right)+H\left(W\right)
}
}D\left(P_{YZ|SX}\|\pi_{YZ|SX}|\pi_{SX}\right).\label{eq:-18-1-1-2}
\end{equation}
\fi
%\color{magenta}
\begin{equation}
\Gamma\left(\pi_{SXYZ},P_{W}\right)\leq\overline{\Delta}\left(\pi_{SXYZ},P_{W}\right):=\inf_{\substack{P_{U},P_{A|SU},P_{B|XU},P_{Y|ABU},P_{Z|ABU}:\\
H\left(U\right) < H\left(ABU|SXYZ\right)+H\left(W\right)
}
}D\left(P_{YZ|SX}\|\pi_{YZ|SX}|\pi_{SX}\right).\label{eq:-18-1-1-2}
\end{equation}
%\color{black}
\iffalse
\color{red}
VA comment: I changed the weak inequality in the constraint to strict inequality
and replaced $\min$ by $\inf$.
\color{black}
\blue
(Lei: Thanks for your revision. Your revision is correct.)
\black
\fi 

Let $\left(Q_{U},Q_{A|SU},Q_{B|XU},Q_{Y|ABU},Q_{Z|ABU}\right)$ be
%a tuple of optimal distributions that attains the minimum in
%\color{magenta}
any tuple of joint distributions that satisfy the constraints on the right hand side of
%\color{black}
\eqref{eq:-18-1-1-2}.
Both Alice and Bob adopt 
%the code 
%\color{magenta}
a coding scheme as
%\color{black}
in the point-to-point setting.
%by using the same codebooks. 
%\blue 
Specifically, 
for the
first  block,
Alice  
%constantly 
sends a sequence of  i.i.d. uniform r.v.'s 
$A_t \sim \mathrm{Unif}(\mathcal{A})$ to Bob, and Bob  
%constantly 
sends a sequence of  i.i.d. uniform r.v.'s 
$B_t \sim \mathrm{Unif}(\mathcal{B})$ to Alice, where $\mathbf{A}_1,\mathbf{B}_1$ are   independent of %$\mathbf{X}_1,\mathbf{Y}_1$. 
%\color{magenta}
$\mathbf{S}_1,\mathbf{X}_1$. 
%\color{black}
\iffalse
Alice generates
$\mathbf{Y}_1$ with a fixed distribution $\hat{Q}_{Y}^{N}$, and Bob generates
$\mathbf{Z}_1$ with a fixed distribution $\hat{Q}_{Z}^{N}$  where $\hat{Q}_{Y},\hat{Q}_{Z}$ are respectively 
  optimal distributions attaining $\Delta_1:=\min_{Q_{Y}}D\left(Q_{Y}\|\pi_{Y|S}|\pi_{S}\right)$ and $\Delta_2:=\min_{Q_{Z}}D\left(Q_{Z}\|\pi_{Z|X}|\pi_{X}\right)$. 
   Note that $\Delta_1,\Delta_2$ are finite by assumption.
\fi
%\color{magenta}
Alice generates $\mathbf{Y}_1$ as a constant sequence equal to $y$ and 
Bob generates $\mathbf{Z}_1$ as a constant sequence equal to $z$
where $(y,z)$ are such that $\pi_{YZ|SX}(y,z|s,x) > 0$ for all $(s,x)$
(the existence of such a pair $(y,z)$ was assumed in the statement of the theorem).
Note that $D(\delta_{(y,z)}\|\pi_{YZ|SX}|\pi_{SX})$ is finite, where 
$\delta_{(y,z)}$ denotes the probability distribution concentrated at $(y,z)$.
%\color{black}
\iffalse
\color{red}
VA comment: Changed the preceding sentence since the earlier version seemed to allow for infinite
relative entropy. Please check.
\color{black}
\blue
(Lei: Thanks for your revision. Your revision is correct.)
\black
\fi
  Furthermore, $\mathbf{M}_1,\mathbf{U}_1$ are  set to  be constant.  
  %Obviously, $\mathbf{A}_1,\mathbf{B}_1, \mathbf{S}_1, \mathbf{X}_1, \mathbf{Y}_1, \mathbf{Z}_1$ are independent of  $\mathcal{C}_1,\hat{\mathcal{C}}_1$.
\iffalse
  \color{red}
  VA comment: Deleted the preceding sentence since it seems unnecessary and the notation 
  $\mathcal{C}_1,\hat{\mathcal{C}}_1$ has not been formally defined in this case. Please check.
\color{black} \blue
(Lei: Thanks for your revision. Your revision is better.)
\black
\fi
For $k$-th block with $k\ge 2$, 
\black
Alice and Bob   individually
generate $\mathbf{M}_{k}=\mathbf{M}\left(\mathbf{A}_{k-1},\mathbf{B}_{k-1},\mathbf{W}_{k-1}\right)$
and $\mathbf{U}_{k}=\mathbf{U}\left(\mathbf{M}_{k}\right)$ by using
the common codebooks, the previous communication bits $\mathbf{A}_{k-1},\mathbf{B}_{k-1}$,
and the common randomness $\mathbf{W}_{k-1}$. Then by using $\left(\mathbf{S}_{k},\mathbf{U}_{k}\right)$,
Alice generates $\mathbf{A}_{k}$ according to the product conditional
distribution $Q_{A|SU}^{N}$ and then sends it to Bob. By using $\left(\mathbf{X}_{k},\mathbf{U}_{k}\right)$,
Bob generates $\mathbf{B}_{k}$ according to the product conditional
distribution $Q_{B|XU}^{N}$ and then sends it to Alice. Upon observing
$\left(\mathbf{A}_{k},\mathbf{B}_{k},\mathbf{U}_{k}\right)$ Alice
generates $\mathbf{Y}_{k}$ according to the product conditional distribution
$Q_{Y|ABU}^{N}$. Upon observing $\left(\mathbf{A}_{k},\mathbf{B}_{k},\mathbf{U}_{k}\right)$
Bob generates $\mathbf{Z}_{k}$ according to the product conditional
distribution $Q_{Z|ABU}^{N}$. 

The distribution synthesized by the code above is exactly the one
synthesized by the following code in the point-to-point setting. Consider
a new scenario in which Alice is a sender and Bob is a receiver.
%\blue 
Specifically, 
for the
first  block,
the encoder  
%constantly 
sends a sequence of  i.i.d. uniform r.v.'s 
$(A_t,B_t) \sim \mathrm{Unif}(\mathcal{A}\times \mathcal{B})$ to the decoder, where $\mathbf{A}_1,\mathbf{B}_1$ are   independent of $\mathbf{S}_1,\mathbf{X}_1$. 
The decoder  generates
$\mathbf{Y}_1\sim \hat{Q}_{Y}^{N}$  and  
$\mathbf{Z}_1\sim \hat{Q}_{Z}^{N}$ independently.
For the $k$-th block with $k\ge 2$, 
%\black
as in the
interactive setting above, Alice and Bob can individually generate
$\mathbf{M}_{k}=\mathbf{M}\left(\mathbf{A}_{k-1},\mathbf{B}_{k-1},\mathbf{W}_{k-1}\right)$
and $\mathbf{U}_{k}=\mathbf{U}\left(\mathbf{M}_{k}\right)$ by using
the common codebooks, the previous communication bits $\mathbf{A}_{k-1},\mathbf{B}_{k-1}$,
and the common randomness $\mathbf{W}_{k-1}$. Alice observes $\left(\mathbf{S}_{k},\mathbf{X}_{k}\right)$,
generates bits $\left(\mathbf{A}_{k},\mathbf{B}_{k}\right)$ according
to the distribution $Q_{A|SU}^{N}Q_{B|XU}^{N}$, and then sends these
bits to Bob. After receiving these bits, Bob generates $\left(\mathbf{Y}_{k},\mathbf{Z}_{k}\right)$
according to the product conditional distribution $Q_{Y|ABU}^{N}Q_{Z|ABU}^{N}$.
 By 
%\color{magenta}
the achievability part of the proof of 
%\color{black}
Theorem \ref{thm:sequentialCS}, the KL divergence induced by
this code is 
%exactly $\overline{\Delta}\left(\pi_{SXYZ},P_{W}\right)$,
%\color{magenta}
bounded above by the term in the infimum on the RHS of \eqref{eq:-18-1-1-2} 
corresponding to the choice that was made of 
$\left(Q_{U},Q_{A|SU},Q_{B|XU},Q_{Y|ABU},Q_{Z|ABU}\right)$. This proves 
\eqref{eq:-18-1-1-2} .
%\color{black}
%and hence, so is the one induced by the interactive code above. 

The random variable $V$ can be added into the optimization  
in the definition of  $\overline{\Delta}\left(\pi_{SXYZ},P_{W}\right)$ by the argument 
given at the end of achievability proof of Theorem \ref{thm:sequentialCS}.
%\color{magenta}
This shows that $\Gamma(\pi_{SXYZ},P_W) \le \Delta^+(\pi_{SXYZ},P_W)$, where
\[
\Delta^{+}\left(\pi_{SXYZ},P_{W}\right):=\inf_{\substack{P_{UV},P_{A|SUV},P_{B|XUV},P_{Y|ABUV},P_{Z|ABUV}:\\
H\left(U|V\right) < H\left(ABU|SXYZV\right)+H\left(W\right)
}
}D\left(P_{YZ|SXV}\|\pi_{YZ|SX}|\pi_{SX}P_{V}\right).
\]
But under the assumption that there is some 
$(y,z)$ such that $\pi_{YZ|SX}(y,z|s,x) > 0$ for all $(s,x)$
one can show by an argument similar to that of 
Lemma \ref{lem:psilem} that $\Delta^{+}\left(\pi_{SXYZ},P_{W}\right) = 
\Delta\left(\pi_{SXYZ},P_{W}\right)$.
%\color{black}
%which coincides the quantity $\Delta\left(\pi_{SXYZ},P_{W}\right)$.

\subsection{Converse}

We next consider the converse part. Observe that 
\begin{align*}
D\left(P_{Y^{n}Z^{n}|S^{n}X^{n}}\|\pi_{YZ|SX}^{n}|\pi_{SX}^{n}\right) & =\sum_{k=1}^{n}D\left(P_{Y_{k}Z_{k}|S^{k}X^{k}Y^{k-1}Z^{k-1}}\|\pi_{YZ|SX}|\pi_{SX}^{k}P_{Y^{k-1}Z^{k-1}|S^{k}X^{k}}\right)
\end{align*}
Denote $K\sim\mathrm{Unif}\left[1:n\right]$ as a random time index,
which is independent of all other r.v.'s involved in the system. Define
$U:=\left(A^{K-1},B^{K-1},W^{K}\right),V:=\left(S^{K-1},X^{K-1},Y^{K-1},Z^{K-1},K\right),A:=A_{K},B:=B_{K},S:=S_{K},X:=X_{K},Y:=Y_{K},Z:=Z_{K}$.
Then 
\begin{align*}
\frac{1}{n}D\left(P_{Y^{n}Z^{n}|S^{n}X^{n}}\|\pi_{YZ|SX}^{n}|\pi_{SX}^{n}\right) & =D\left(P_{YZ|SXV}\|\pi_{YZ|SX}|\pi_{SX}P_{V}\right)
\end{align*}
It is easy to verify that 
\iffalse
\begin{align*}
P_{UVABSXYZ} & =P_{K}(k)P_{W}^{k}\pi_{SX}^{k-1}P_{A^{k-1}|S^{k-1}W^{k-1}}P_{B^{k-1}|X^{k-1}W^{k-1}}P_{Y^{k-1}|A^{k-1}B^{k-1}S^{k-1}W^{k-1}}P_{Z^{k-1}|A^{k-1}B^{k-1}X^{k-1}W^{k-1}}\\
 & \qquad\times\pi_{SX}P_{A_{k}|S^{k}A^{k-1}B^{k-1}W^{k}}P_{B_{k}|X^{k}A^{k-1}B^{k-1}W^{k}}P_{Y_{k}|A^{k}B^{k}S^{k}Y^{k-1}W^{k}}P_{Z_{k}|A^{k}B^{k}X^{k}Z^{k-1}W^{k}}\\
 & =P_{UV}\pi_{SX}P_{A|SUV}P_{B|XUV}P_{Y|ABSUV}P_{Z|ABXUV}
\end{align*}
\fi
%\color{magenta}
\begin{align*}
P_{UVABSXYZ} & =P_{K}(k)P_{W}^{k}\pi_{SX}^{k-1}P_{A^{k-1}B^{k-1}Y^{k-1}Z^{k-1}|S^{k-1}X^{k-1}W^{k-1}}\\
 & \qquad\times\pi_{SX}P_{A_{k}|S^{k}A^{k-1}B^{k-1}Y^{k-1}W^{k}}P_{B_{k}|X^{k}A^{k-1}B^{k-1}Z^{k-1}W^{k}}P_{Y_{k}|A^{k}B^{k}S^{k}Y^{k-1}W^{k}}P_{Z_{k}|A^{k}B^{k}X^{k}Z^{k-1}W^{k}}\\
 & =P_{UV}\pi_{SX}P_{A|SUV}P_{B|XUV}P_{Y|ABSUV}P_{Z|ABXUV}.
\end{align*}
% P_{Y^{k-1}|A^{k-1}B^{k-1}S^{k-1}W^{k-1}}P_{Z^{k-1}|A^{k-1}B^{k-1}X^{k-1}W^{k-1}}
%\color{black}
Hence it remains to show $H\left(U|V\right)\leq H\left(ABU|SXYZV\right)+H\left(W\right)$.
This can be easily verified similarly to \eqref{eq:-38}-\eqref{eq:-39}.

\iffalse
\color{red}
VA question: In the long displayed equation above, I changed $P_{A^{k-1}|S^{k-1}W^{k-1}}P_{B^{k-1}|X^{k-1}W^{k-1}}$ to 
$P_{A^{k-1}B^{k-1}|S^{k-1}X^{k-1}W^{k-1}}$.
This seems to be important and does not seem to affect the proof. Please check.
\color{black}
\blue
(Lei: Thanks for your revision. Your revision is better.)
\black
\fi

 \bibliographystyle{unsrt}
\bibliography{ref}

\end{document}

%% file: preamble.tex
\usepackage[mathscr]{eucal}
\usepackage{fixltx2e}%ordering of single and double column floats
\usepackage{array}%array and tabular environments
\usepackage{verbatim}
\usepackage{bm}
\usepackage{algorithmic}
\usepackage{algorithm}
\usepackage{verbatim}
\usepackage{textcomp}
\usepackage{mathrsfs}
\usepackage{epstopdf}

\newcommand{\openone}{\leavevmode\hbox{\small1\normalsize\kern-.33em1}}

\catcode`~=11 \def\UrlSpecials{\do\~{\kern -.15em\lower .7ex\hbox{~}\kern .04em}} \catcode`~=13

\allowdisplaybreaks[4]

\newcommand{\calM}{\mathcal{M}}

\newcommand{\calW}{\mathcal{W}}
\newcommand{\calX}{\mathcal{X}}

\DeclareMathAlphabet{\mathbsf}{OT1}{cmss}{bx}{n}
\DeclareMathAlphabet{\mathssf}{OT1}{cmss}{m}{sl}% slanted sans serif

\DeclareSymbolFont{bsfletters}{OT1}{cmss}{bx}{n}
\DeclareSymbolFont{ssfletters}{OT1}{cmss}{m}{n}
\DeclareMathSymbol{\bsfGamma}{0}{bsfletters}{'000}
\DeclareMathSymbol{\ssfGamma}{0}{ssfletters}{'000}
\DeclareMathSymbol{\bsfDelta}{0}{bsfletters}{'001}
\DeclareMathSymbol{\ssfDelta}{0}{ssfletters}{'001}
\DeclareMathSymbol{\bsfTheta}{0}{bsfletters}{'002}
\DeclareMathSymbol{\ssfTheta}{0}{ssfletters}{'002}
\DeclareMathSymbol{\bsfLambda}{0}{bsfletters}{'003}
\DeclareMathSymbol{\ssfLambda}{0}{ssfletters}{'003}
\DeclareMathSymbol{\bsfXi}{0}{bsfletters}{'004}
\DeclareMathSymbol{\ssfXi}{0}{ssfletters}{'004}
\DeclareMathSymbol{\bsfPi}{0}{bsfletters}{'005}
\DeclareMathSymbol{\ssfPi}{0}{ssfletters}{'005}
\DeclareMathSymbol{\bsfSigma}{0}{bsfletters}{'006}
\DeclareMathSymbol{\ssfSigma}{0}{ssfletters}{'006}
\DeclareMathSymbol{\bsfUpsilon}{0}{bsfletters}{'007}
\DeclareMathSymbol{\ssfUpsilon}{0}{ssfletters}{'007}
\DeclareMathSymbol{\bsfPhi}{0}{bsfletters}{'010}
\DeclareMathSymbol{\ssfPhi}{0}{ssfletters}{'010}
\DeclareMathSymbol{\bsfPsi}{0}{bsfletters}{'011}
\DeclareMathSymbol{\ssfPsi}{0}{ssfletters}{'011}
\DeclareMathSymbol{\bsfOmega}{0}{bsfletters}{'012}
\DeclareMathSymbol{\ssfOmega}{0}{ssfletters}{'012}

\DeclareMathOperator{\supp}{supp}